\documentclass[12pt,semidraft]{article}

\usepackage{array} 
\usepackage{epsfig}
\usepackage{amssymb}
\usepackage{amsmath}                          
\usepackage{graphics,graphpap}
\usepackage{graphicx}
\usepackage{epstopdf}

\renewcommand{\thefootnote}{\fnsymbol{footnote}}

\def\fslash#1{\setbox0=\hbox{$#1$}%
\rlap{\ifdim\wd0>.7em\kern.22\wd0\else\kern.1\wd0\fi /}#1}

\def\simge{\mathrel{%
   \rlap{\raise 0.511ex \hbox{$>$}}{\lower 0.511ex \hbox{$\sim$}}}}

\def\simle{\mathrel{
   \rlap{\raise 0.511ex \hbox{$<$}}{\lower 0.511ex \hbox{$\sim$}}}}

\newcommand{\al}{\alpha}

\newcommand{\matel}[3]{\langle #1|#2|#3\rangle}

\newcommand{\msm}[1]{ m_{#1}}
\newcommand{\mph}[1]{ {{m}_{#1,\rm ph}}  }
\newcommand{\mphq}[1]{ {{m}^{2}_{#1,\rm ph}}  }
\newcommand{\Tanh}[1]{ \tanh({ #1}) }
\newcommand{\Cosh}[1]{ \cosh({ #1}) }
\newcommand{\Sinh}[1]{ \sinh({ #1}) }

\begin{document}


\begin{titlepage}
\begin{flushright}\begin{tabular}{l}
IPPP/08/21 \\
DCPT/08/42
\end{tabular}
\end{flushright}
\vskip1.5cm
\begin{center}
   {\Large \bf \boldmath Electroweak Precision Data  \\[0.1cm] 
  and  the Lee-Wick Standard Model}
    \vskip1.3cm {\sc \large
 Thomas~E.~J.~Underwood\footnote{Thomas.E.J.Underwood@mpi-hd.mpg.de}$^{1}$ and  Roman~Zwicky\footnote{Roman.Zwicky@durham.ac.uk}$^{2}$}
   \vskip0.3cm
 $^1$ Max-Planck-Institut f\"ur Kernphysik, Saupfercheckweg 1, 69117
     Heidelberg, Germany 
     \vskip0.1cm
         $^2$ IPPP, Department of Physics, 
Durham University, Durham DH1 3LE, UK \\
\vskip1.5cm 
\vskip0.5cm

{\large\bf Abstract:\\[8pt]} \parbox[t]{\textwidth}{ We investigate the
  electroweak precision constraints on the recently proposed Lee-Wick Standard
  Model at tree level.  We analyze low energy, Z-pole (LEP1/SLC) and LEP2 data
  separately. We derive the exact tree level low energy and Z-pole effective
  Lagrangians from both the auxiliary field and higher derivative formulation
  of the theory.  For the LEP2 data we use the fact that the Lee-Wick Standard
  Model belongs to the class of models that 
  assumes a so-called`universal' form which can be described by
  seven oblique parameters at leading order in $\msm{W}^2/M_{1,2}^2$.   At tree
  level we find that $Y = -
  \msm{W}^2/M_1^2$ and $W= - \msm{W}^2/M_2^2$, where the negative sign is due
  to the presence of the negative norm states. 
  All other oblique parameters $(\hat S,X)$ and 
  $(\hat T,\hat U,V)$ are found to be zero.
 In a separate addendum we show how our results differ from previous investigations, where contact terms, which are found to be of leading order, have been neglected.
The LEP1/SLC constraints are slightly stronger than LEP2 and much stronger than
  the low energy ones.  The LEP1/SLC results exclude gauge boson masses of
  $M_1 \simeq M_2 \sim 3\, {\rm TeV}$ at the $99\%$ confidence level.
  Somewhat lower masses are possible when one of the masses assumes a large
  value.  Loop corrections to the electroweak observables are suppressed by
  the standard $\sim 1/(4\pi)^2$ factor and are therefore not expected to change the constraints
 on $M1$ and $M_2$.  This assertion is most transparent from
  the higher derivative formulation of the theory. \\
  }
\vfill

\end{center}
\end{titlepage}

\setcounter{footnote}{0}
\renewcommand{\thefootnote}{\arabic{footnote}}

\newpage

\newpage

\section{Introduction}

In 1970 Lee and Wick (LW) proposed a finite theory of QED
\cite{negmetric,finiteqed}. Just over a year ago, Grinstein, O'Connell
and Wise (GOW) \cite{GOW} extended those ideas to non-Abelian gauge theories
and chiral fermions and constructed a Lee-Wick Standard  Model (LWSM).
The essence of the work by LW was to introduce Pauli-Villars, wrong-sign
propagator, fields as physical degrees of freedom. Since the interaction
terms only contain certain combinations of fields and ghost fields, the
latter can be integrated out at the cost of introducing higher derivative (HD)
interactions. The new degrees of freedom lead to amplitudes which are better
behaved in the ultraviolet and render the logarithmically divergent QED
finite.  It was shown by GOW \cite{GOW} that the LWSM is free from quadratic
divergences and therefore provides a possible solution to the `hierarchy
problem'. It was also shown that the addition of very heavy right handed neutrinos, in the context of the see-saw mechanism, does not
destabilize the Higgs mass \cite{rnu}.

The introduction of the Pauli-Villars wrong sign states
brings into question basic concepts such as  causality and unitarity.
These issues were investigated
in some detail in the 1970's and some results are summarized
in the Erice Lectures of Lee \cite{Lee-Erice} and
Coleman \cite{Coleman-Erice}. In reference \cite{Coleman-Erice} it is
illustrated that the wrong sign of the width
compensates for the wrong sign in the propagator to
ensure unitarity in a simple s-channel process. The wrong sign of the width in turn implies that poles will move into the physical sheet which demands a new contour prescription \cite{S-gang}.
The modification of the contour 
might bring into question Lorentz invariance, 
c.f.\,references \cite{Nakanishi}
for criticism and  \cite{response} for a response.
In summary, the current status is that
there are no known examples in perturbation theory which are in conflict
with unitarity when applying the contour modification of \cite{S-gang}
and acausal phenomena arises only at scales which are not accessible to current
experiments. The phenomenon of causality was
reconsidered only very recently in an $O(N)$ model \cite{O(N)}.
The authors investigate the large N limit, where
the theory is described by a single one-loop bubble,
and obtain a unitary  and Lorentz invariant scattering
amplitude.
Furthermore, a non-perturbative definition through the path integral
via the contour deformation \cite{S-gang} does
not seem to be straightforward or conclusive \cite{BoulawareGross}.
A higher derivative version of the LW Higgs sector was used
for  lattice field theory \cite{Jansen_formal,Jansen_lattice}. This
is of great interest because it can smooth cut-off dependences.
As stated in reference  \cite{Jansen_formal}
this does not really clarify the issue in Minkowski space since an
analytic continuation is prevented by the complex ghost poles.
A path integral approach with test functions was
recently proposed \cite{tonder}, from where the contour prescription can be derived.

Following the proposal of the LWSM several phenomenological investigations
have been pursued.  The low Higgs mass discovery channel $gg \to h_0 \to
\gamma \gamma$ was found to be moderately positively enhanced, $5$-$20\%$ for
the top LW at $0.5$-$1\,{\rm TeV}$ \cite{KUZ}. It is a curious fact, or a
rather unique signature of the ghost fields, that the CKM elements $|V_{\rm
  tx}|$ are accompanied by an enhancement factor $(1+ (m_t/M_{\rm LW})^2/2)$
which can lead to $|V_{\rm tx}| \gtrsim 1$ at the few percent level
\cite{KUZ}.  Flavour changing neutral currents induced by integrating out
heavy LW fermions have been found to give acceptably small contributions for a
LW mass scale $M_{\rm LW} \sim 1 \,{\rm TeV} \cite{DW}$.  LW gauge bosons were
found to lead to possible signatures of a unique nature both at the LHC in
dijet channels \cite{RizzoLHC} and in cross-sections and left-right
asymmetries in Bhabba scattering at linear colliders \cite{RizzoILC}.

Aspects of LW gauge theories unrelated to the LWSM have also found attention.
The running of a non-Abelian LW theory coupled to a scalar field was
investigated in \cite{beta}, where it was found that the gauge coupling runs
faster than in an ordinary gauge theory.  It was also shown that massive LW
gauge bosons do not require a Higgs degree of freedom to unitarize amplitudes
at high energy \cite{massive}. This is because the formulation is gauge
invariant without a Higgs field and corresponding Ward identities assure a
moderate growth at high energy.  Abelian and non-Abelian LW gauge theories
have also been to shown to give rise to chiral symmetry breaking \cite{Xsim}.
Moreover it has been suggested that gravitational 1-loop corrections may lead
to the appearance of a Lee-Wick photon \cite{grav-phot}. This effect has
recently also been studied in the context of large extra dimensions
\cite{grav-phot2}.

In this paper we analyse the constraints on the LWSM gauge boson sector coming
from low energy, $Z$-pole and LEP-2 data.  We will work at tree level, an
approximation justified as we do not see any reason for loop corrections to
compete with tree level contributions relieving the $\sim 1/(4\pi)^2$
hierarchy.  Even though we find that among the three symmetry classes of the
oblique parameters two are vanishing at tree level, it is not necessary to
calculate the loop contributions to those parameters as the seven oblique
parameters all have similar experimental constraints and are typically of the
same order when no symmetry is protecting them. Possibly an exception to this
rule is the breaking of custodial symmetry at loop level due to the mass
splitting of the third family. This contributes to the rho parameter as
$\Delta \rho_*(0) = \alpha T = \hat T$.  It is well known that the
dominant correction to the rho parameter in the SM is due to fermion loops and
is given by $\Delta \rho_*(0)^{\rm SM} \simeq G_F m_t^2/( 8 \pi^2 \sqrt{2})
\simeq 10^{-2}$ e.g. \cite{PT}.  An easy way to obtain a crude estimate of
this contribution in LWSM is to take a look at the HD formulation,
\begin{equation}
\label{eq:form}
\hat \Delta(p) = \frac{1}{p^2 - p^4/M_{\rm LW}^2 - m^2}  \,, \qquad
\hat S_F(p) = \frac{\fslash{p}\big(1-\frac{p^2}{M_{\rm LW}^2}\big) +
m}{p^2(1-\frac{p^2}{M_{\rm LW}^2}\big)^2 -m^2} \, .
\end{equation}
indicating a contribution to the rho parameter $|\Delta \rho_*(0)^{\rm LWSM}| \sim {\cal O}(1)\,10^{-2}
(m_t/M_{\rm LW})^2$, up to further 
logarithmic corrections.  From the  constraint
 $|\Delta \rho_*(0)| < 10^{-3}$ \cite{PDG} one 
 could then deduce limits on  $m_{\tilde t}$.
 
We feel obliged to comment on other papers that have investigated electroweak
 precision constraints on the LWSM \cite{EW1,lebed}. 
Our results differ conceptually  
from theirs by including effects of contact terms.
The latter are found to be of leading order and
are therefore necessary ingredient to a consistent
analysis. Further details about this relation can be 
found in a separate addendum to this paper.
  We would like to mention that our results for the oblique
 parameters, which rely on an expansion in $\msm{W,Z}^2/M_{\rm new}^2$ are
 backed-up by an exact treatment of the tree level low-energy and $Z$-pole
 Lagrangians. Further comments and more details on the origin of these
 discrepancies can be found in an addendum at the end of this paper.

 We would also like to make some comments on observables which we have omitted
 in this paper. For instance one of the LWSM contributions to the anomalous
 magnetic moment of the muon, $(g-2)_\mu$ is given by Schwinger's vertex
 correction where the photon is replaced by the Lee-Wick photon. The form of
 the propagators in the LWSM suggests that this contribution is suppressed
 with respect to the corresponding SM contribution by a factor $(m_\mu/M_1)^2
 \sim 10^{-8}$, with $M_1 \sim 1 \,{\rm TeV}$ according to the high energy
 constraints investigated in this paper.  This value is about two orders of
 magnitude too low to explain the difference between theory and experiment.
 Notice that the situation is rather different to the MSSM, for example,
 because the contraints on the scale analogous to $M_1$ are not so tight there
 since tree level constraints from other precision data are absent. The MSSM
 contribution is also sensitive to enhancements by factors of $\tan \beta$.
 The contributions to the rate for $b \to s \gamma$ would be interesting to
 study. The dominant contribution is due to penguins with top loops and we
 would therefore expect to either obtain constraints on the LW fermion mass
 scale or better agreement between experiment and theory.

We  refrain from reviewing the LWSM itself in more detail and refer instead to the original paper \cite{GOW}
and our own paper in connection with the matrix
notation for the LW generations \cite{KUZ}.
The paper is organized as follows. In section \ref{sec:EWC} we analyse
the electroweak sector of the LWSM. We derive the low energy effective Lagrangian
at tree level in the auxiliary field and the HD formulation in subsection
\ref{sec:low}. The effective Lagrangian relevant to data collected at the $Z$-pole is derived in section
\ref{sec:Z-pole} in both formalisms. The LEP-2 data is assessed
via the oblique approximation in section \ref{sec:LEP2}, where we also rederive the results of the previous section in the oblique approximation.
In section \ref{sec:fit} we relate the parameters of the effective
Lagrangian to the observables and detail the sources of our
experimental input and procedures. Concluding remarks can be found in section \ref{sec:con}.

Details of the gauge
boson propagators in the higher derivative formalism are given
in appendix \ref{app:WZprop}. Some exact results for
diagonalization of the gauge boson mass matrices, for the
case where the SU(2)$_L$ and U(1)$_Y$ LW extensions
are degenerate in their mass scale $M_1 = M_2$,
are given in  appendix \ref{app:m12eq}.

Throughout the paper we neglect terms of the order of
$m_f m_f'/M_{\rm LW}^2$, where the $f$ is any fermion
other than the top quark. In particular, this implies that we
are allowed to  omit contributions from longitudinal components of gauge bosons ${\cal O}(p_\mu p_\nu)$ and neglect diagonalization of mass matrices in the fermion sector (although this was outlined in \cite{KUZ}).

\section{Effective Lagrangians for electroweak constraints}
\label{sec:EWC}

The low energy effective Lagrangian, Z-pole observables and the constraints
from LEP-2 are investigated in sections \ref{sec:low}, \ref{sec:Z-pole},
\ref{sec:LEP2} respectively. The low energy Lagrangian is rather
straightforward whereas the exact Z-pole effective Lagrangian demands the
diagonalization of the Z-boson sector.  In the LEP-2 section we will exploit
the fact that the LWSM belongs to the class of ``universal'' Lagrangians which
allow the parameterization of the leading electroweak corrections in terms of
seven oblique parameters $(S,T,U,V,W,X,Y)$.  By leading we mean to first order
in $\msm{W}^2/M^2$ where $M$ is the mass scale of the LW gauge bosons. We will
rederive all the results of the previous sections in this approximation.  The
reader who is familiar with electroweak precision data and the formalisms used
to constrain it can directly go to section \ref{sec:LEP2} with the additional
information that among the three best measured parameters \cite{PDG},
\begin{eqnarray}
\label{eq:best}
 \mph{Z} &=& 91.1876(21)\,{\rm GeV} \,, \nonumber \\
  \alpha(\mph{Z}) &=& \frac{1}{127.918(18)}   \,, \nonumber \\
  G_F &=&  1.16637(1) \cdot  10^{-5} \,{\rm GeV^2} \, ,
\end{eqnarray}
only $\mph{Z}$ receives corrections. This implies a correction to the Weinberg
angle.

\subsection{Low energy effective Lagrangian}
\label{sec:low}

The low energy effective electroweak Lagrangian can be parametrised as
\begin{equation}
\label{eq:para}
{\cal L}^{\rm eff}_{\rm EW} =
-\frac{4 G_F}{\sqrt{2}}
\Big( J_+ \!\cdot\! J_- + \rho_*(0)  J_{\rm nc}^2  \Big) + C_Q J_Q^2\,,
\end{equation}
with
\begin{equation}
\label{eq:curr}
J_{\pm}^\mu = (J_1 \pm i J_2)^\mu \,, \qquad    J_{\rm nc}^\mu \equiv
(J_3 -s^2_*(0) J_Q)^\mu \, .
\end{equation}
in terms of four parameters
\begin{equation}
\label{eq:plow}
p_{\rm low} \equiv [\rho_*(0), s^2_*(0),G_F , C_Q]\,,
\end{equation}
which  assume the values
\begin{equation}
\label{eq:plowRES}
(p_{\rm low})_{SM}  = [1, \sin(\theta_W)^2,\frac{1}{\sqrt{2} v^2} ,0]\,,
\end{equation}
in the SM.
In the following two subsections we will derive the values of the four low
energy parameters   in
the LWSM using both the auxiliary field and the higher derivative (HD) formalisms.

\subsubsection{Auxiliary field formalism}

The low energy effective Lagrangian can be derived in the auxilliary field
picture by integrating out the heavy gauge degrees of freedom (i.e. all gauge
fields except for the photon). To do this in an efficient manner  
the matrix formalism introduced in reference \cite{KUZ} is extended
to the gauge boson current sector,
\footnote{Contributions from unphysical Higgs bosons and terms of the $O(p_\mu
  p_\nu)$ are suppressed at least by a factor of $m_f m_{f'}/m_W^2$ and we
  shall neglect them here and thereafter in connection with the low energy
  observables.}
\begin{equation}
{\cal L}_{\rm EW} = - {\cal J}^{{\rm a}\,\top} \!\cdot\! {\cal W}^{{\rm
a}\,} - {\cal J}^{0\,\top}  \!\cdot\!  {\cal B}
+ \frac{1}{2}\,{\cal W}^{{\rm a}\,\mu\,\top} {\cal M}_{\cal
W}\,\eta_{2}\,{\cal W}^{{\rm
    a}}_{\mu} + \frac{1}{2}\,{\cal B}^{\mu\,\top} {\cal M}_{\cal
B}\,\eta_4\,{\cal B}_{\mu} + \ldots\,,
\end{equation}
with
\begin{alignat}{2}
{\cal J}^{{\rm a}\,\top}_{\mu} & =  g_2 \big( J,\,  J \big)^{\rm a}_{\mu} \,, \qquad 
& {\cal J}^{0\,\top}_{\mu}  = & \,\, \big(g_1 J^{Y},\, g_2 J^{3},\, g_1
  J^{Y},\, g_2 J^{3}\big)_{\mu} \,,\nonumber\\
{\cal W}_\mu^{{\rm a}\,\top} & =  \big(W,\, \widetilde{W} \big)^{\rm
    a}_{\mu} \,,
& {\cal B}_\mu^\top  = & \,\, \big(B ,\, W^3 ,\, \tilde{B} ,\,
\widetilde{W}^3 \big)_\mu\,,
\end{alignat}
where $J^{\rm a}_{\mu}$ and $J^{Y}_{\mu}$ are the appropriate fermion
currents, ${\cal M}_{\cal B}$ and ${\cal M}_{\cal W}$ are the neutral and
charged gauge boson mass matrices, $\eta_2 = {\rm diag}(1,-1)$, $\eta_4 =
{\rm
  diag}(1,1,-1,-1)$ and ${\rm a} = \{1,2\}$. The dots refer to cubic  and
quartic
couplings between the gauge bosons. For ${\cal M}_{\cal B}$ and ${\cal
  M}_{\cal W}$ we have
\begin{eqnarray}
\label{eq:M}
{\cal M}_{\cal B}\,\eta_4 & = &
\left( \begin{array}{lc}
M_{\rm SM}\hspace*{5mm} & M_{\rm SM}\\
M_{\rm SM}\hspace*{5mm} & M_{\rm SM} - M_{12} \end{array} \right)\,,
\quad
M_{\rm SM} \ =\ \frac{v^2}{4} \left( \begin{array}{cc}
g_1^2 & -g_1\,g_2\\
-g_1\,g_2 & g_2^2 \end{array} \right)\,,\nonumber\\[5pt]
{\cal M}_{\cal W}\,\eta_2 & = &
\frac{1}{4} \left( \begin{array}{cc}
g_2^2\,v^2\hspace*{5mm} & g_2^2\,v^2\\
g_2^2\,v^2\hspace*{5mm} & g_2^2\,v^2 - 4 M_2^2\end{array} \right)\,,\quad
M_{12} \ =\ \left( \begin{array}{cc}
M_1^2 & 0\\
0 & M_2^2 \end{array} \right)\,.
\end{eqnarray}

The {\bf charged current sector} is straightforward. Integrating out the
charged gauge bosons at tree level is equivalent to applying the equation of
motions (eom)
\begin{equation}
\label{eq:eomW}
{\cal W}^{\rm a}_\mu  = 
\big({\cal M}_{\cal W}\,\eta_2\big)^{-1}\, {\cal J}^{\rm a}_\mu\,,
\end{equation}
for the ${\cal W}^{\rm a}_\mu$ fields.

The {\bf neutral sector} is more involved because 
the massless photon has to be decoupled in 
order to invert the mass matrix.  
This is easily done using the transformation ${\cal B} \to
S\,{\cal B}^\prime$ where $S\,\eta_4\,S^\dagger\,\eta_4 = \mathbb{I}$ and
\begin{equation}
\label{eq:weinrot}
S = \left( \begin{array}{cc}
R_W & 0_2 \\
0_2   & \mathbb{I}_2
\end{array}\right) \qquad  {\rm with} \qquad 
R_W = \left( \begin{array}{cc} 
c & -s \\
s   & c
\end{array}\right) \,,
\end{equation}
and 
\begin{equation}
\label{eq:sc}
s \equiv \sin \theta_W \,, \qquad c \equiv \cos \theta_W \,, \qquad 
t \equiv \tan \theta_W \equiv
g_1/g_2 \,.
\end{equation}
It is straightforward to verify that $S^\top\,{\cal M}_{\cal B}
\eta_4\,S$ is
block-diagonal with one zero eigenvalue.

Denoting the projection of the primed currents and mass matrices on the heavy
gauge boson subspace by a double prime, the neutral gauge boson Lagrangian
reads
\begin{equation}
\label{eq:decoupledphoton}
{\cal L} =  - {\cal J}^{0 \prime\prime\,\,\top} \!\cdot\!
{\cal B}^{\prime\prime} + \frac{1}{2}\,{\cal
B}^{\prime\prime\,\mu\,\top} {\cal M}^{\prime\prime}_{\cal
B}\,\eta_3\,{\cal B}^{\prime\prime}_{\mu}  \,\, -e\,A\!\cdot\! J^Q\,,
\end{equation}
where $e\equiv g_2 s$,
\begin{equation}
{\cal B}^{\prime\prime\top}_\mu = \big(c W^3 - s B, \tilde{B},
\widetilde{W}^3\big)_\mu\,,\qquad
{\cal J}^{0 \prime\prime\top}_\mu = e \left( \frac{c}{s}
J^{3} - \frac{s }{c} J^{Y}, \frac{1}{c} J^{Y},\frac{1}{s}
J^{3} \right)_\mu \,,
\end{equation}
and
\begin{equation}
\label{eq:defmb}
{\cal M}^{\prime\prime}_{\cal B}\,\eta_{3} = \msm{Z}^2 \,\left(
\begin{array}{ccc}
1 & -s & c\\
-s & \;\;\;s^2 - x_1 & -s\,c\\
c & -s\,c & c^2 - x_2
\end{array}\right)\,,
\end{equation}
where $\eta_3 = {\rm diag} (1,-1,-1)$ and
\begin{equation}
\label{eq:defxi}
 x_i \equiv \frac{M_i^2}{\msm{Z}^2}\,, \qquad  
\msm{Z}^2 \equiv \frac{e^2 v^2}{4 s^2 c^2}\,,
\end{equation}
are notations frequently used throughout the paper.

The massive neutral gauge bosons may be integrated out by substituting the
following expression obtained from the eom for ${\cal B}^{\prime\prime}_{\mu}$
into Eq.~(\ref{eq:decoupledphoton}), i.e.
\begin{equation}
{\cal B}^{\prime\prime}_{\mu} = \big({\cal M}_{\cal
B}^{\prime\prime}\,\eta_3\big)^{-1}
{\cal J}^{0 \prime\prime}_\mu \, .
\end{equation}
After some algebra, the electroweak low-energy effective Lagrangian can
now be written down
\begin{equation}
{\cal L}^{\rm eff}_{\rm EW} = -\ e A \!\cdot\! J^Q \ -\ \frac{2}{v^2} \Big(
J_+ \!\cdot\! J_- \ +\  \big(J_3 -
s^2 J_Q \big)^2 \Big) \ +\  \frac{e^2}{2}\Big(\frac{c^2}{M_1^2} + 
\frac{s^2}{M_2^2}\Big)  J_Q^{2}\,.
\label{eq:lowenergy}
\end{equation}
By comparing Eq.~(\ref{eq:lowenergy}) with Eq.~(\ref{eq:para}) we can read off
expressions for the  the Fermi constant
$G_F$ and the parameters $\rho_*(0)$, $s^2_*(0)$ and $C_Q$
\footnote{In reference \cite{GOW} a contribution to the rho parameter 
was obtained in the approximation of retaining only the mixed terms,
$A_{\rm SM} A_{\rm LW}$, in the mass matrix.} 
\begin{eqnarray}
\label{eq:low}
G_F = \frac{1}{\sqrt{2}\,v^2}\,, \quad
\rho_*(0) = 1\,, \quad
\quad s^2_*(0) = s^2\,, \quad C_Q = \frac{e^2}{2}\Big(\frac{c^2}{M_1^2} + 
\frac{s^2}{M_2^2}\Big) \,.
\end{eqnarray}
Note that the electromagnetic coupling $4 \pi \alpha = e^2$ is not affected
since the photon cannot be integrated out. The coefficient $C_Q$ cannot 
be seen at low energies anticipating the LW gauge boson masses to be around  
$M_i \sim 1\,{\rm TeV}$,
because it is shielded by the photon background  by a very small
factor $s_{\rm c.m.}/M_i^2$. The scale $s_{\rm c.m.}$ refers to data points in 
$e^+ e^- \to {\rm hadrons}$ \cite{PDG} 
and $s_{\rm c.m.} \ll \msm{Z}^2$ as otherwise the
effective description breaks down.
The $C_Q$ term corresponds 
to a contact term originating from the massive LW photon.
We already want to point out at  this stage that the low energy observables
will receive corrections through $s$ due to a shift in $\mph{Z}$, 
which we will derive in the next chapter.

\subsubsection{Higher derivative formalism}
\label{sec:HDlow}

In order to derive the low energy effective Lagrangian in the higher
derivative picture we need the $W$ and $Z$ propagators in this formalism.  The
coupling of the gauge bosons is identical 
to the SM, up to
corrections of the type $\partial^2/M_{LW}^2$, which are irrelevant at low
energy.  The SM gauge boson current Lagrangian is given by
\begin{equation}
\label{eq:current-boson}
{\cal L} =  - \frac{g_2}{\sqrt{2}} (W^+   \!\cdot\! J^+ +  W^-   \!\cdot\! J^- ) -
\frac{g_2}{c} Z  \!\cdot\! J_{\rm nc}
- e A  \!\cdot\!  J_Q \,  ,
\end{equation}
where $W^{\pm} = (W^1 \pm i W^2)/\sqrt{2}$ and the currents have been defined in 
Eq.~\eqref{eq:curr}.
It is rather straightforward to show that the HD propagator assumes
the following form
\begin{equation}
\label{eq:WHD}
\hat D^W_{\mu\nu}(p^2) = \hat D^W(p^2) \Big( - g_{\mu\nu} + p_{\mu} p_\nu
f^W_{pp} \Big) \,, \quad \quad
D^W(p^2) = \frac{i}{p^2 - \msm{W}^2 -p^4/M_2^2} ,
\end{equation}
with 
\begin{equation}
\label{eq:mwsm}
\msm{W}^2 \equiv  \frac{e^2 v^2}{4 s^2}\,,
\end{equation}
in analogy with the definitions \eqref{eq:defxi}.
 More details and an explicit expression for $f_{pp}^W$
are given in appendix \ref{app:WZprop}.
The charged low energy effective action follows then from Eq.~\eqref{eq:current-boson}
\begin{equation}
\label{eq:chargedHD}
{\cal L}^{\rm eff}_{\rm CC} =   - i \hat D^W(0) \frac{g_2^2}{2} J_+ \!
\cdot \! J_- =
 -  \frac{g_2^2}{2 m_W^2} J_+ \! \cdot \! J_-  \,.
\end{equation}
The propagator of the neutral gauge bosons
$\hat N = (\hat Z,\hat A)$ has the form
\begin{equation}
\hat D_{\mu \nu}^{N}(p^2) = \hat D^{N}(p^2) (- g_{\mu \nu} + p_\mu p_\nu
f^{N}_{pp} )\,.
\end{equation}
This propagator is non-diagonal in the $(\hat Z,\hat A)$ space if $M_1
\neq M_2$. Further details can be found in appendix
\ref{app:WZprop} Eq.~\eqref{eq:DNexp}
The neutral low energy effective Lagrangian is given by
\begin{eqnarray}
\label{eq:neutralHD}
{\cal L}_{\rm NC}^{\rm eff} &=&
-i \frac{g_2^2}{2 c^2} \Big(   J_N^{\mu \,\,T} (\hat D^N(p^2) -
\left( \begin{array}{cc}
0 \hspace*{5mm} & 0 \\
0 \hspace*{5mm} & 1 \end{array} \right)
 \frac{i}{p^2} ) J_N^\mu  \Big)_{p^2 \to 0} \,,\\[0.1cm]
&=& -\frac{g_2^2}{2 m_W^2} J_Z^2 +  \frac{e^2}{2} \Big(  \frac{c^2}{M_1^2} +
\frac{s^2}{M_2^2}  \Big) J_Q^2\,,
\end{eqnarray}
with the photon pole subtracted since the photon cannot be integrated out.  We
have implicitly used the notation $J_N = (\frac{g_2}{c} J_Z, e J_Q )$.  From
the low energy effective interactions (\ref{eq:chargedHD}) and
(\ref{eq:neutralHD}) and the parametrisation (\ref{eq:para}) and
(\ref{eq:curr}) we read off the same parameters as in Eq.~\eqref{eq:low}.

\subsection{Effective Lagrangian at the Z-pole}
\label{sec:Z-pole}

When considering experimental data collected at LEP and SLC around the
$Z$-pole, generic ``new physics'' can be parameterised by the Lagrangian
\begin{equation}
\label{eq:split}
{\cal L} = {\cal L}^{SM}(e,s,v) + \delta {\cal L}^{\rm new}(e,s,v,..)\,.
\end{equation}
However, the parameters fitted to experimental data are not the ($e,s,v$) but
other parameters ($e_0,s_0,v_0$), which are defined using the three best
measured observables $\alpha$, $G_F$ and $\mph{Z}$ mentioned
at the beginning of this section. 
These may be written
\begin{alignat}{2}
\label{eq:00}
m_{Z,0}^2 &\equiv \mphq{Z} = \msm{Z}^2( 1 + \delta_Z)   
\qquad \qquad \qquad  & & \msm{Z} = \frac{e v}{2cs} \nonumber \\
\alpha_0(\mph{Z}) &\equiv \alpha(\mph{Z})  = \alpha(\mph{Z})_{SM}(1 + \delta_\alpha) 
& & \alpha(\mph{Z})_{SM} = \frac{e^2}{4 \pi} \nonumber \\
G_{F,0} &\equiv G_F =  (G_F)_{SM}(1+\delta_G) 
& & (G_F)_{SM} =  \frac{1}{\sqrt{2} v^2}
\end{alignat}

Using the measured values of $G_F$, $\alpha$ and $\mph{Z}$ we can obtain the
following three fundamental parameters of the SM Lagrangian
\begin{eqnarray}
\label{eq:famous}
& & e = \sqrt{\frac{4 \pi \alpha(\mph{Z})}{1+\delta_\alpha}}\,, \qquad 
v  = \sqrt{ \frac{(1+\delta_G)}{\sqrt{2} G_F}}\,, \nonumber \\
& & s^2 c^2 = s_0^2 c_0^2 \frac{(1+\delta_G)(1+\delta_Z)}{1+\delta_\alpha}\,, 
\end{eqnarray}
with the well-measured intermediate quantity $s_0$ defined as
\begin{equation}
\label{eq:celebre}
\frac{1}{4} \sin^2(2 \theta_0) = s_0^2 c_0^2 \equiv \frac{\pi \alpha(\mph{Z})}{\sqrt{2}
G_F M_Z^2}\, ,
\end{equation}
with $s_0 = \sin(\theta_0)$ and $c_0 = \cos(\theta_0)$. As mentioned in the
introduction, the low energy observables will receive corrections due to the
non-trivial relation between $(s,c)$ and $(s_0,c_0)$ given in
Eq.~\eqref{eq:famous}.

It is convenient to parameterise couplings of
the $Z$ to fermions in terms of the 
following generalised Lagrangian \cite{burgess}
\begin{equation}
{\cal L}_Z^{\rm eff} = -\,\frac{e_0}{s_0 c_0} \sum_i \bar{f}_i \gamma^\mu \Big[ \big(
g^{f,SM}_L + \delta g^{f}_L \big) P_L + \big(
g^{f,SM}_R + \delta g^{f}_R \big) P_R \Big] f_i\,Z_\mu\,,
\label{eq:effZpole}
\end{equation}
where $P_{L,R}$ are the usual left and right projection operators and
\begin{equation}
g^{f,SM}_L = t_3^f - q^f s_0^2\,,\qquad g^{f,SM}_R = - q^f s_0^2\,,
\end{equation}
are the tree-level SM couplings.  Corrections then arise through
\vspace{0.2cm}
\begin{enumerate}
\item  the new interactions in $\delta {\cal L}^{\rm new}(e,s,v,..)$ from Eq.~\eqref{eq:split},
\item on the parameters $(e,s,v)$ via Eq.~\eqref{eq:famous} due to the presence of
  $\delta {\cal L}^{\rm new}(e,s,v,..)$.
\end{enumerate}
\vspace{0.2cm}
For comparison with other work, we write the effective Lagrangian \begin{equation}
\label{eq:Leff}
{\cal L}_Z^{\rm eff} = - \left( \rho_f \sqrt{2} G_F \right)^{1/2}\,2\,
\mph{Z} \,J_Z \! \cdot \!  Z \,, \qquad J_Z^\mu = J_3^\mu - s_*^2 (\mph{Z})\,J_Q^
{\mu}\,,
\end{equation}
in an alternative notation with the intermediate quantities $\rho_f$ and
$s_*(\mph{Z})$. The changes in the $Z$ pole couplings \eqref{eq:effZpole} in
terms of these variables are
\begin{eqnarray}
\label{eq:dg}
\delta g^{f}_L = t_3^f (\sqrt{\rho_f} -1) - q^f(\sqrt{\rho_f} s_*^2(\mph{Z})-s_0^2)\,,
\qquad 
\delta g^{f}_R =  - q^f(\sqrt{\rho_f} s_*^2(\mph{Z})-s_0^2)
\,.
\end{eqnarray}
In the following two subsections we will derive the expressions for the three
$Z$-pole parameters, $p_{Z\,pole} \equiv [\rho_f,s^2_*(\mph {Z}),\mph{Z}]$ and
the $W$-boson mass, $\mph{W}$ in the auxiliary field and HD formalism.

\subsubsection{Auxiliary field formalism}

In the auxiliary field picture, an effective Lagrangian of the form
\eqref{eq:effZpole} can be derived by integrating out all the heavy neutral
gauge bosons apart from the $Z$. This may be accomplished by block
diagonalising the mass matrix ${\cal M}^{\prime\prime}_{\cal B}$ defined in
Eq.~(\ref{eq:defmb}). We find it convenient to  use  the following ansatz 
\begin{equation}
\label{eq:Q}
Q \equiv 
\left(\begin{array}{cc}
1 & 0_{1\times2}\\
0_{2\times1} & R_W
\end{array}\right)
\left(\begin{array}{ccc}
\Cosh{\phi} & \Sinh{\phi} & 0\\
\Sinh{\phi} & \Cosh{\phi} & 0\\
0 & 0 & 1
\end{array} \right)
\left(\begin{array}{ccc}
\Cosh{\theta} & 0 & \Sinh{\theta}\\
0 & 1 & 0\\
\Sinh{\theta} & 0 & \Cosh{\theta}
\end{array} \right)
\,,
\end{equation}
which acts as ${\cal B}^{\prime\prime} \to Q {\cal B}^{\prime\prime\prime}$
with $Q \eta_3 Q^\dagger \eta_3 = \mathbb{I}$ as usual. The conditions for
block diagonalisation then give two equations which can be used to relate
$\phi$ and $\theta$ to $M_1$ and $M_2$.  In terms of $x_i$ defined in
Eq.~\eqref{eq:defxi} we find
\begin{eqnarray}
\label{eq:x12}
x_1 &=& \Big(1+ \frac{\Tanh{\theta}}{\Cosh{\phi}}\Big)\big(1+ f(c,s,\theta,\phi) \big)\,, 
\nonumber  \\
x_2 &=& \Big(1+ \frac{\Tanh{\theta}}{\Cosh{\phi}}\Big)\big(1+ f(s,-c,\theta,\phi) \big)
\,,
\end{eqnarray}
with 
\begin{equation}
f(c,s,\theta,\phi)  = \frac{s\,\Cosh{\theta} \Cosh{\phi} }{s\,\Sinh{\theta}- c\,\Cosh
{\theta} \Sinh{\phi}}\,.
\end{equation}
Notice that the limit $M_1 = M_2$ corresponds to $\Sinh{\phi} = 0$ (see also
appendix \ref{app:m12eq}).

For the sake of clarity, let us state here that the combined action of the
transformations $S$ and $Q$, defined in Eqs.~\eqref{eq:weinrot} and
\eqref{eq:Q} respectively, leads to an overall transformation of
\begin{equation}
 S_{\rm tot} = S 
\left( \begin{array}{ll}
1 & 0_{3 \times 1} \\
0_{1 \times 3} & Q
\end{array}\right)\,,
\end{equation}
on ${\cal M_B}\,\eta_4$ such that
\begin{equation}
 S_{\rm tot}^T {\cal M_B}\eta_4  S_{\rm tot} = 
 \left( \begin{array}{ll}
{\cal M}_{AZ,\rm ph}  & 0_2 \\
0_2 &  - {\cal M}_{\tilde A\tilde Z} 
\end{array}\right)
\, ,   \qquad {\cal M}_{AZ,\rm ph}  =
 \left( \begin{array}{ll}
0 & 0 \\
0 &  \mphq{Z} 
\end{array}\right) \,,
\end{equation}
where $ {\cal M}_{\tilde A \tilde Z} $ is the mass matrix for the heavy Lee-Wick gauge
bosons $(\tilde A,\tilde Z)$ which we do not need to diagonalize to obtain the
$Z$-pole Lagrangian.  Nevertheless, in appendix \ref{app:m12eq} we have
diagonalised ${\cal M_B}$ completely for the special case $M_1 = M_2$ ($\Sinh 
\phi = 0$).

Now, after applying the transformation ${\cal B}^{\prime\prime} \to Q {\cal
  B}^{\prime\prime\prime}$ to the Lagrangian in Eq.~\eqref{eq:decoupledphoton},
the SM-like $Z$ boson can be decoupled from the other neutral heavy LW bosons
and the effective neutral current Lagrangian takes the form
\begin{eqnarray}
{\cal L}_{\rm nc} & = & - {\cal J}^{0 \prime\prime \,\top}   \!\cdot\! Q {\cal
  B}^{\prime\prime\prime} + \frac{1}{2}\,{\cal
  B}^{\prime\prime\prime\,\top}_\mu {\cal M}^{\prime\prime\prime}_{\cal
  B}\,\eta_3\,{\cal B}^{\prime\prime\prime}_{\mu} 
 \ =\  {\cal L}_Z^{\rm eff} + \frac{1}{2}\,\mphq{Z}
Z^2  + \ldots
\end{eqnarray}
where ${\cal L}_Z^{\rm eff}$ is defined in Eq.~\eqref{eq:Leff} and
 the three Z-pole parameters 
are given by 
\begin{alignat}{2}
\mphq{Z}  &= \msm{Z}^2(1 + \delta_Z) \,,\qquad  \quad 
& \delta_Z  =& \frac{\Tanh{\theta}}{\Cosh{\phi}}\,,
\nonumber \\
\rho_f &= \cosh^2(\theta) \,\cosh^2(\phi) \left(1 + \delta_Z\right)\,,  &\phantom
{m} &  \nonumber \\
\label{eq:smz}
s_*^2(\mph{Z}) &= s^2 \left(1+\delta_s\right)\,,\qquad & \delta_s =& -\frac{c}{s}\,
\frac{\Tanh{\phi}}{1+\delta_Z}\,.
\end{alignat}
We would like to emphasize here that $\rho_f \geq 1$ can be
independently\footnote{ From Eq.~\eqref{eq:smz} 
$\rho_f \geq 1$ with  $\delta_Z \geq 1$ which follows from $x_1(x_2) \geq 1 + \delta_Z$ which is a
  very mild assumption.}  inferred from the low energy effective Lagrangian.
Since $\rho_*(0) = 1$ obtained in \eqref{eq:low} is composed of $\rho_*(0) =
\rho_f - \rho_f^{\tilde A} - \rho_f^{\tilde Z}$, where $\rho_f^{\tilde
  A(\tilde Z)} \geq 0$ are the analogue of $\rho_f$ ($\rho_f= \rho_f^Z$ in
this notation) for an $\tilde A(\tilde Z)$ pole effective Lagrangian and the
minus sign is due to the ghost nature of these gauge boson states.  Notice
that from the low-energy effective Lagrangian we have
\begin{equation}
\label{eq:dadG}
\delta_\alpha = 0\,,\qquad \delta_G = 0\,,
\end{equation}
for the parameters defined in Eq.~\eqref{eq:00}.


Finally, although not strictly a $Z$-pole observable, the $W$-boson mass can
be derived from the matrix ${\cal M}_{\cal W}$. From reference \cite{KUZ}
\begin{equation}
\label{eq:mw}
\mphq{W} = \msm{W}^2 
\frac{1}{2}(x_{W} - \sqrt{x_{W}^2 - 4 x_{W}}) \,, \qquad x_{W} = \frac{M_2^2}{\msm
{W}^2}\,.
\end{equation}
The $\mphq{W}$ mass can then be expressed in terms of the 
angles $(\phi,\theta)$ via $x_W = x_2/c^2$ in Eq.~\eqref{eq:x12}.

Notice that in the limit of zero Weinberg angle $\theta_W \to 0$ (i.e. $s \to
0, c\to 1$), which is the limit of exact custodial symmetry SU(2)$_V$ at tree
level, the physical $W$ and $Z$-boson masses unite,
\begin{equation}
\label{eq:trip}
\mphq{Z(W)} \stackrel{\theta_W \to 0}{\to} \msm{W}^2 ( 1 + \Tanh{\theta})
\end{equation}
to form  the custodial  SU(2)$_V$  isotriplet.

In section \ref{sec:fit} we will use the expressions (\ref{eq:dg}) with
(\ref{eq:smz}) and (\ref{eq:mw}) to derive the corrections to the electroweak
precision observables. Notice that the parameter $s$ will be expressed in
terms of the measured value $s_0$ according to Eq.~\eqref{eq:famous}.

\subsubsection{Higher derivative formalism}
\label{sec:zpoleHD} 
 
In this subsection we will show how to derive the parameters $\rho_f$
[Eq.~\eqref{eq:smz}] and $s_*^2(\mph{Z})$ [Eq.~\eqref{eq:smz}] from the HD
formalism. We do not discuss the determination of $\mph{Z,W}$ from the
viewpoint of the HD formalism as they do not lead to further insight.  These
parameters are reveiled as multiplicative factors to the HD propagator $\hat
D^N(p^2)$ \eqref{eq:DN}.  Identifying with the parametrisation of the Z-pole
effective Lagrangian \eqref{eq:Leff}
\begin{eqnarray}
\label{eq:trick}
\lim_{p^2 \to \mphq{Z}} (-i) \Big[J_N^{\;T} \hat D^N(p^2)(p^2-\mphq{Z}) J_N\Big]
= (J_3-s_*^2 J_Q)^2 \rho_f  \underbrace{(2\mph{Z})^2 \sqrt{2} G_F}_{\frac{g_2^2}
{c^2}(1+\delta_Z)}
\end{eqnarray}
with the previously used notation  $J_N^{\;T} = (\frac{g_2}{c} J_Z,e J_Q)$
and the limits
\begin{eqnarray*} 
\lim  \limits_{p^2 \to\mphq{Z}}  (p^2 -  \mphq{Z}) (\hat D^N)_{XY}(p^2)
=
\left\{ 
\begin  {array}  {ll}   \def  \ma  { \left(  \begin {array} {ccccc} } 
  [\Cosh{\phi} \Cosh{\theta}(1 + \delta_Z)]^2 & \quad XY = ZZ \\
  \sinh^2(\phi) \cosh^2(\theta)  &  \quad  XY = AA \\
  \Cosh{\phi} \cosh^2(\theta) \Sinh{\phi}(1 + \delta_Z)  &   \quad XY = ZA 
\end{array}  \right.
\quad .
\end{eqnarray*}
we obtain
\begin{eqnarray}
\rho_f &=& \cosh^2(\theta) \cosh^2(\phi) (1 + \delta_Z)\,, \nonumber  \\
s^2_*(\mph{Z}) &=&  s^2 ( 1 + \delta_s)\,,  \qquad 
\end{eqnarray}
in accordance with Eq.~\eqref{eq:smz}. The Z boson mass, or
$\delta_Z$ \eqref{eq:smz}, is given by the lowest root of the polynomial
in the denominator of the propagator $(\hat D^N)$ as implicitly used in 
the equation above.

The case $M_1 = M_2$ i.e. $\Sinh{\phi} =0$ is discussed in appendix \ref
{app:HDM} is very instructive since it can be discussed analytically from
where it is easily understood that $s_*^2(\mph{Z}) \to s^2$ when $\Sinh{\phi}
\to 0$ for instance.

\subsection{LEP-2 data and the oblique parameters}
\label{sec:LEP2}

At LEP-2 cross sections of the type $\sigma( e^+e^- \to f \bar f)$ and
forward-backward asymmetries $A^f_{ FB}$ were measured for centre of mass
energies in the range $130$-$209\,{\rm GeV}$, around the Z-pole \cite{LEP2}.
The observables are the same as those used in Z-pole measurements which are
summarized in appendix \ref{app:zpoleobs}. The LEP-2 measurements allow
constraints to be set on contact or current-current terms,
\begin{equation}
\label{eq:contact}
{\cal L}^{\rm eff} = c_f \, J_f^2  \,.
\end{equation}
In the LWSM, as in many other models, such contact terms arise from
integrating out heavy gauge bosons. The current-current terms are of dimension
six and it is possible to incorporate these effects with an effective field
theory to that order.

Since the LWSM belongs to the so-called universal class of models
\cite{ItalySpain}, its effective field theory is described by the so-called
`oblique' parameters. This description incorporates corrections due to new
physics to leading order in\footnote{Reference \cite{BeyondSTU} nicely
  describes how to extend the formalism to the case when the new physics is
  close to the $M_Z$-scale.}
\begin{equation}
\label{eq:eps}
\epsilon = \frac{\msm{W,Z}^2}{M_{\rm new}^2}\,.
\end{equation}
It can be shown that corrections to the Z-pole observables and measurements of
$\sigma( e^+e^- \to f \bar f)$ at LEP-2 \cite{ItalySpain} and to a great
extent corrections to the low energy observables \cite{strumia06} can be
written as a set of seven parameters which are straightforward to calculate
and do not necessitate the diagonalization of the $Z$ and $W$ boson mass
matrices.

A model is said to be universal in this context if its effective theory at the
$M_Z$ scale is described by an effective Lagrangian of the type:
\begin{equation}
\label{eq:universal}
{\cal L}^{\rm universal} = \frac{1}{2}  A^a_\mu \,
\Pi^{ab}(q^2)\, A_\mu^b + g A^a \! \cdot \! J_{\rm SM}^a \,,
\end{equation}
where we have used $q^2$ instead of the partial derivative for notational
simplicity.  The longitudinal part ${\cal O}(q_\mu q_\nu)$ is omitted since it
is suppressed by a factor of ${\cal O}(m_f^2/M_W^2)$ as mentioned previously.
The gauge index $a$ runs over the electroweak gauge sector SU(2)$_L
\times$U(1)$_Y$ $a = (1,2,3,B)$.  As the notation suggests $J_{SM}^a$ are the
SM currents.
The fields  $A^a$  on the other hand couple to the SM states, 
as $\matel{0}{A^a}{A_{\rm SM}^b} \sim \delta^{ab}$
in the sense of
interpolating fields, but are in general different from 
the SM fields. The Lagrangian \eqref{eq:universal} essentially
corresponds to the SM Lagrangian with self energy corrections and to
non-diagonal gauge fields. The latter influences predictions only at
${\cal O}(\epsilon^2)$ and therefore the interpolating fields are sufficient.

Exploiting the assumed hierarchy \eqref{eq:eps} the $\Pi(q^2)$ function can be
expanded
\begin{equation}
\label{eq:expand}
\Pi(q^2) = \Pi(0) + \Pi'(0)q^2 + \frac{1}{2}  \Pi''(0) q^4  + {\cal O}
(q^6) \, ,
\end{equation}
where the expansion has to be carried out to ${\cal O}(q^4)$ in order to
consistently account for the contact terms \eqref{eq:contact}.  There are
twelve parameters corresponding to all possible combinations of $\{
\Pi_{ab}(0), \Pi_{ab}'(0) , \Pi''_{ab}(0)\}$ 
with $ ab \in \{BB,B3,33,11\} $.
Two are zero due to $U(1)_{Q}$ gauge invariance or the masslessness of the
photon and three are absorbed into the definition of the three best measured
electroweak parameters listed in Eq.~\eqref{eq:best}, leaving a total of seven
parameters.  As emphasized in \cite{ItalySpain} these seven parameters fall
into three classes according to (custodial, SU(2)$_L$) symmetry\footnote{ In the limit  
$g_1 \to 0$ the SM has a global
SU(2)$_L\times$SU(2)$_R$ symmetry in the 
absence of a Higgs VEV. When the latter arises 
the symmetry breaks down to its diagonal subgroup
SU(2)$_V$ \cite{Veltman}. Extending this notion
to the case $g_1 \neq 0$ is in principle ambiguous.
In the case where there are no additional weak 
gauge bosons, such as for example technicolor, it is 
sensible to extend the notion to be the symmetry 
that  maintains $m_W^2/(m_Z^2 \cos^2(\theta_W)) = 1$ \cite{SSVZ}. It was termed 
custodial symmetry in honor of its protective function. When there are additional gauge bosons it is not
clear how to extend this notion. We adopt here the
classification of reference \cite{ItalySpain}. Under this notation custodial symmetry could 
for instance mean that the low energy rho parameter 
$\rho_*(0)$ \eqref{eq:para}, 
which is the ration of charged to neutral currents, 
remains unity c.f. section \ref{sec:reder}. 
Importantly in the limit 
$g_1 \to 0$ an SU(2) symmetry is recovered 
in the LWSM, e.g. degeneracy of the weak
gauge boson masses  Eq.~\eqref{eq:trip}. 
We will therefore refrain in this paper from 
using the term custodial symmetry other than
for this classification.}.  The first class
$(+,-)$ violates only SU(2)$_L$ symmetry\footnote{
In the modern literature, e.g. \cite{PDG}, the oblique parameters 
solely contain contributions from physics 
beyond the (minimal) SM and therefore
a value for the Higgs mass has to assumed
for the SM predictions.
In earlier times  contribution from the Higgs and the top were sometimes also absorbed into the oblique parameters \cite{PT}.}
\begin{equation}
\hat S =  \frac{g_2}{g_1} \Pi'_{3B}(0), \, \qquad X =  
\frac{\msm{W}^2}{2}  \Pi''_{3B}(0) \, .
\end{equation}
The second class $(-,-)$ violates both symmetries
\begin{eqnarray}
\hat T &=& \frac{1}{\msm{W}^2}(  \Pi_{33}(0) -   \Pi_{11}(0)) \,, \qquad 
 \hat U = -(\Pi'_{33}(0) - \Pi'_{11}(0))\,, \nonumber \\
 V &=& \frac{\msm{W}^2}{2} (  \Pi''_{33}(0) -   \Pi''_{11}(0))  \,.
\end{eqnarray}
The third class $(+,+)$ does not violate any of those symmetries
\begin{equation}
Y = \frac{\msm{W}^2}{2}    \Pi''_{BB}(0) \,, \qquad W  = \frac{\msm{W}^2}{2}    \Pi''_{33}(0) \, .
\end{equation}
There is no fourth $(-,+)$ class since a violation of custodial symmetry in this
context also implies a violation of SU(2)$_L$.  Note that for an expansion up
to the first order in $ \Pi (q^2)$ only the variables $\hat S, \hat T$ and
$\hat U$ are required which correspond to the three oblique parameters used
for Z-pole physics, c.f. \cite{PT} and references therein, up to normalisation
factors.

The fact that the LWSM corresponds to the class of universal Lagrangians is
most easily recognized in the HD formulation\footnote{Of course the auxiliary
  field (AF) formulation can also be brought into a universal form.  As emphasized
  in \cite{ItalySpain,strumia06} for instance the class of universal theories
  is somewhat larger than usually thought of. The criterion is that only gauge
  bosons with SU(2)$_L \times$U(1)$_Y$ quantum numbers couple to the SM
  currents. One then integrates out the linear combinations of heavy gauge
  bosons which do not couple to the SM currents,
  which is   in the LWSM, 
  in order to bring the
  effective Lagrangian into a 'universal' form. In
  the AF formulation of the LWSM it is 
  $\tilde A$-$A$ 
  which does not
  couple to the SM currents and integrating out those degrees of freedom then
  exactly reproduces the HD formulation.}  of the theory which assumes a
universal form \eqref{eq:universal} with
\begin{eqnarray}
\label{eq:LWPI}
\Pi^{ab}(q^2) &=& \mathbb{I}_2 q^2- M_{\rm SM} -   q^4 M_{12}^{-1} 
\nonumber \\
\Pi^{11}(q^2)    &=&   q^2 - \msm{W}^2  -  \frac{q^4}{M_2^2} 
\end{eqnarray}
where $a,b \in \{B,3\}$ and $M_{\rm SM}$ and $M_{12}$ are
defined in Eq.~\eqref{eq:M}. 

Since the LWSM neither violates SU(2)$_L$ nor custodial symmetry in the gauge
boson sector the first class $(+,-)$
\begin{equation}
\hat S = 0 \,, \qquad X = 0\,,
\end{equation}
and the second class $(-,-)$ 
\begin{equation}
\label{eq:Class2}
\hat T = 0 \, \qquad \hat U =0  \,, \qquad V = 0\,,
\end{equation}
are identically zero. The only non-vanishing values are
found in the third class $(+,+)$ 
\begin{equation}
\label{eq:WY}
Y = - \frac{\msm{W}^2}{M_1^2} = - \frac{c^2}{x_1} \,, \qquad W = - \frac{\msm{W}^2}{M_2^2} = - \frac{c^2}{x_2} \,,
\end{equation}
which does not violate the symmetries. 

In the following three subsections we will first rederive the results of the
previous sections in terms of the oblique parameters, comment on the sign of
$W$ and $Y$ and point towards a gluonic constraint testable at the LHC. These
three subsections can be omitted for the reader interested in the constraints
on $M_{1,2}$ only.

\subsubsection{Low energy and Z-pole  results in terms of oblique parameters \\
at leading order in $\msm{W}^2/M_{LW}^2$. }
\label{sec:reder}

In this subsection we will generally not distinguish between $s(c)$ and
$s_0(c_0)$ because this is a next-to-leading order effect except when we
derive the leading order difference between $s^2$ and $s_0^2$.  In order to
rederive the results of sections \ref{sec:low} and \ref{sec:Z-pole} we have to
express the results directly in terms of $M_{1,2}^2$ at leading order. We have
obtained almost all the results in these sections in terms of $\delta_Z =
\Tanh{\phi}/\Cosh{\theta}$, $\delta_s = - c/s \Tanh{\phi} /(1+\delta_Z)$ where
$(\phi,\theta)$ are the hyperbolic rotation angles linked to the LW mass
scales $M_1$ and $M_2$ via Eq.~\eqref{eq:x12}.  We may rewrite the latter
system of equations as
\begin{eqnarray}
\label{eq:x12new}
\frac{1+\delta_Z}{x_1 -(1+\delta_Z)} &=& \delta_Z -\frac{1}{t} \tanh(\phi)\,,
\nonumber \\
\frac{1+\delta_Z}{x_2 -(1+\delta_Z)} &=& \delta_Z + t  \tanh(\phi)\,,
\end{eqnarray}
from which follows
\begin{equation}
\tanh(\phi) = \frac{1+\delta_Z}{t+1/t}\left( \frac{1}{x_2 -(1+\delta_Z)}-
\frac{1}{x_1 -(1+\delta_Z)}\right)  \, .
\end{equation}
A simple or leading order solution $\delta_Z$ is obtained when the $x_i -
(1+\delta_Z)$ is replaced by $x_i -1$ in the denominator of the system
Eq.~\eqref{eq:x12new}
\begin{equation}
\delta_Z = \frac{ (x_2 -1)t^2 +(x_1-1)}{(x_2 -1)(x_1-2)t^2 +(x_1 -1)(x_2-2)} + 
{\cal O}\left(\frac{1}{x_{1,2}^{2}}\right) .
\end{equation}
Expanding in inverse powers of $x_i$ with 
\begin{equation}
x_1,x_2 \gg 1 \,,\qquad \frac{1}{x_1}-\frac{1}{x_2} \sim {\cal O}(1)\,,
\end{equation}
we obtain, with $\delta_s$ from Eq.~\eqref{eq:smz},
\begin{eqnarray}
\label{eq:approx}
\delta_Z &=& \frac{c^2}{x_2} + \frac{s^2}{x_1} + {\cal O}\left(\frac{1}{x_{1,2}^{2}}\right)\,,  \nonumber\\ 
\delta_s &=& c^2\left(\frac{1}{x_1} - \frac{1}{x_2} \right) \left[ 1 + \left(\frac{1}{x_2} + 
    \frac{1}{x_1}\right)\delta_Z \right] + \ldots  = c^2 \left(\frac{1}{x_1} - \frac{1}{x_2} \right) +
  {\cal O}\left(\frac{1}{x_{1,2}^{2}}\right)\,.
\end{eqnarray}

The {\bf low energy} data $p_{\rm low} \equiv [\rho_*(0), s^2_*(0),G_F ,
C_Q]$, Eq.~\eqref{eq:plow}, of a universal theory can be found by transforming the physical 
parameters expressed in term of the correlation
functions, e.g., into the
set of seven oblique parameters
\begin{eqnarray}
\rho_*(0) &=& \frac1 {1 - \hat  T} \simeq  1 + \hat T = 1\,, \nonumber \\
s^2_*(0) &=& s^2 = s_0^2\left[1  + \frac{1}{c^2-s^2} \left(\hat S - c^2(\hat T + W) - 
s^2 Y + 2 sc \, X \right) \right]\,, \nonumber \\
& =&  s_0^2  +  \frac{c^2 s^2 }{c^2-s^2}\left(\frac{c^2}{x_2}+ \frac{s^2}{x_1}
\right)\,.
\label{eq:formofds}
\end{eqnarray} 
We do not present expressions for $G_F$ and $C_Q$ as they do not lead to
further insight and parallel the derivation in subsection \ref{sec:HDlow}. In
particular we choose to use $G_F$ as an input parameter, c.f.
Eq.\eqref{eq:best}. The parameters $\rho_*(0)$ and $s_*^2(0)$ do indeed
correspond to the results found in Eq.~\eqref{eq:low} when taking the
linearization of Eq.~\eqref{eq:famous} into account, $s^2 \simeq s_0^2 + c^2
s^2/ (c^2-s^2)\,\delta_Z$, with $\delta_Z$ found using Eq.~\eqref{eq:approx}. The second formula is
also given in reference \cite{ItalySpain}.

The {\bf Z-pole} data, $p_{Z\,pole} \equiv [\rho_f,s^2_*(\mph {Z}),\mph{Z}]$
and $\mph{W}$, can be written in terms of oblique parameters using the
expressions in reference \cite{PT}, 
\begin{eqnarray}
\label{eq:first}
  s^2_*(\mph{Z }) &=& s^2_*(0) + \frac{s}{c}\left[
  (c^2-s^2) X + sc(W - Y) \right] = s^2_*(0) + s^2 c^2 \left( \frac{1}{x_1} - \frac{1}{x_2} \right)\,,
  \nonumber \\
 \frac{ \mphq{W}}{\mphq{Z}}- c_0^2 
 &=& \frac{c^2}{c^2-s^2}\Big(c^2 \hat T - (c^2-s^2)(\hat U -V) - 2(s^2 \hat S + cs X) + s^2 (W + Y)\Big) 
 \nonumber \\
 &=& \frac{-c^2 s^2}{c^2-s^2}\Big(\frac{c^2}{x_1} +
 \frac{c^2}{x_2}\Big)  \,.
\end{eqnarray}
We have again inserted the expressions for the oblique parameters in the LWSM
after the second equality sign. The expression for $s^2_*(\mph{Z })$ is
correct bearing in mind that $s_*^2(0) = s^2$, Eq.~\eqref{eq:low}, and that
$s^2_*(\mph{Z }) = s^2 (1+ \delta_s)$ with $\delta_s$ from
Eq.~\eqref{eq:approx}. The second equation
corresponds to the Veltman rho parameter 
and the formula is a generalization of an expression
given in \cite{PT}. It's verification in the 
context of the LWSM follows from,
\begin{equation*}
\label{eq:second}
 \frac{ \mphq{W}}{\mphq{Z}}- c_0^2  \simeq (s_0^2-s^2) + c^2 (\delta_W-\delta_Z) = \frac{- c^2 s^2}{c^2-s^2} \delta_Z +  c^2 (\delta_W-\delta_Z) = \frac{-c^2 s^2}{c^2-s^2}\Big(\frac{c^2}{x_1} +
 \frac{c^2}{x_2}\Big) \,,
\end{equation*}
with $\delta_W = c^2/x_2 $ 
defined as $\mphq{W} \simeq \msm{W}^2(1+\delta_W)$ in Eq.~\eqref{eq:mw} and 
the expression for $\delta_Z$ in Eq.~\eqref{eq:approx}.

\subsubsection{On the (negative) sign of $W$ and $Y$}
\label{sec:sign}
It was pointed out in reference \cite{strumia06} that the two point functions
of the gauge bosons can be written in terms of a K\"all\'en-Lehmann dispersion
relation
\begin{equation}
\frac{1}{\Pi(q^2)} = \int_{\rm cut} ds \frac{\rho(s)}{q^2-s} \,,
\end{equation}
where $\rho(s)$ is the spectral function given by
\begin{equation}
\label{eq:spectral}
\rho_{11}(q^2)\,\theta(q_0)\,\big(- g_{\mu\nu}  + {\cal O}(q_\mu q_\nu)\big) =
 (2\pi)^3  \sum_\alpha \delta^{(4)}(q-p_\alpha) \, \matel{0}{W^1_\mu}{\alpha}
 \matel{\alpha}{W^1_\nu}{0}\,,
\end{equation}
for the case $a,b= 1$, for example. It is then an elementary exercise to show
that
\begin{equation}
\label{eq:sec}
\Pi''(0) = 
\frac{ \int_{\rm cut}  ds_1 ds_2 \rho(s_1)\rho(s_2) (s_1 - s_2)/(s_1^3 s_2^3)}
{[\int_{\rm cut} ds \rho(s)/s]^3} \, .
\end{equation}
Since the form of Eq.~\eqref{eq:spectral} suggests that $\rho(s) \geq 0$, it
is concluded in \cite{strumia06} that $W,Y \geq 0$ from Eq.~\eqref{eq:sec},
which is indeed the case in many models.  It is therefore not surprising that
in the  LWSM, $W,Y \leq 0$ as a consequence of the negative normed states
which contribute with a negative sign to $\rho(s)$.

It was mentioned in reference \cite{strumia06} that when the SM gauge groups
are embedded into a larger group then $Y$ and $W$ could also turn out to be
negative because ghost states could dominate in the non-gauge invariant
$\Pi(q^2)$.

\subsubsection{A gluonic operator constraint}

To order ${\cal O}(q^4)$ in Eq.~\eqref{eq:expand} there is
also a gluonic operator \cite{ItalySpain}
\begin{equation}
Z =  \frac{\msm{W}^2}{2}  \Pi''_{GG}(0)
\end{equation}
which is sensitive to operators of the type $ (D_\al G_{\mu\nu}^a)^2/2$.
This operator is not related to electroweak symmetry breaking and
the constraints looked at in this paper, but it can be tested at the LHC
possibly in dijet channels investigated in \cite{RizzoLHC}.
It is as simple as before to make a leading order prediction in 
the LWSM
\begin{equation}
 Z = - \frac{\msm{W}^2}{M_3^2} \,,
\end{equation}
where $M_3$ is the mass scale of the LW gluon term
\begin{equation}
\delta {\cal  L} = \frac{1}{M_3^2}\,{\rm Tr} 
	\big(\hat{D}^\mu \hat{G}_{\mu \nu}\big)\,
	\big(\hat{D}^\lambda \hat{G}_\lambda{}^\nu \big) \, .
\end{equation}

\section{The precision observables}
\label{sec:fit}

As discussed in the introduction, throughout the paper we have followed the
usual procedure of dividing the precison observables into 3 classes; low
energy data, data collected in $e^+ e^-$ collisions at the Z-resonance and
data collected in $e^+ e^-$ collisions at LEP-2. In this chapter we present the
numerical constraints on the LW masses $M_1$ and $M_2$ provided by each
data-set.

Predictions for all observables are calculated by splitting each one into a SM
prediction plus a linearised correction due to the LW operators. This
approximation should be valid as long as the corrections are small, which must
be the case since the quality of the SM fit to the data is very good.

To produce the SM predictions we use the 2008 version of the GAPP code
\cite{gapp} with the fixed input parameters in table~\ref{tab:gapp}.

\begin{table}
\begin{center}
\begin{tabular}{cc}
\hline
\hline
Parameter & Value\\
\hline
$\mph{Z}$~[GeV] & $91.1875$\\
$\mph{t}$~[GeV] & $172.6$\\
$\mph{H}$~[GeV] & $115$\\
$\alpha_s$ & $0.120$\\
$\Delta \alpha^{(3)}_{\rm had}$ & $0.00577$\\
$\widehat{m}_c (\mu = \widehat{m}_c)$~[GeV] & $1.290$\\
$\widehat{m}_b (\mu = \widehat{m}_b)$~[GeV] & $4.207$\\
\hline
\hline
\end{tabular}
\end{center}
\caption{\sl Fixed GAPP input parameters used to produce the SM predictions
  used in the $\chi^2$ fits.}
\label{tab:gapp}
\end{table}

\subsection{Low energy}

Precision constraints on the low energy Lagrangian come from several sources.
We utilise results from neutrino-nucleon and neutrino-electron scattering
experiments and measurements of electron-nucleon interactions made by studying
atomic parity violation.  The parameterization of the low energy Lagrangian
differs for each class of experiments and the various parameters used are
defined in appendix~\ref{app:low}. As discussed in the introduction, we
do not include constraints from the anomalous magnetic moment of the muon,
$(g-2){}_\mu$ since we expect the correction in the LWSM to be small compared
to the SM contribution.

The parameters  $\epsilon_{L,R}(q)$, determined by neutrino-nucleon scattering, in terms
of the parameters defined in Eqs.(\ref{eq:para}) and (\ref{eq:curr}), are parametrised as \begin{eqnarray}
  \epsilon_L(q) & = & \rho_*(0)\,\left[t_3^q- q^q\,s^2_*(0)\right]\,,\nonumber\\
  \epsilon_R(q) & = & \rho_*(0)\,\left[- q^q\,s^2_*(0)\right]\,,
\end{eqnarray}
where $t_3^q$ and $q^q$ are respectively the weak isospin and electric charge
of the quark $q$. Experimental determinations of $\epsilon_{L,R}$ are strongly
correlated and so a parameterization in terms of $g_i^2$ and $\theta_i$
($i=L,R$) is often used (see appendix~\ref{app:low}).  In our fits we
use experimental values provided in the 2008 particle data book \cite{PDG},
which are listed in Table~\ref{tab:lowenergy}. Notice that the NuTeV result
($g_L^2$) has been adjusted to take the strange quark asymmetry into account
\cite{erler}.

In the same way, the neutrino-electron scattering parameters $g_{V,A}^{\nu e}$
and the $C_{iq}$ parameterizing electron-nucleon interactions can be all be
written in the form
\begin{eqnarray}
g_{V}^{\nu e} & = & 2\rho_*(0)\,\left[s^2_*(0)-\frac{1}{4}\right]\,,\nonumber\\
g_{A}^{\nu e} & = & -\rho_*(0)/2\,,\nonumber\\
C_{1q} & = & -\rho_*(0) \left[ t_3^q - 2 q^q s_*^2(0)\right]\,,\nonumber\\
C_{2q} & = & -\rho_*(0)\,t_3^q \left[ 1 - 4 s_*^2(0)\right]\,.
\end{eqnarray}
In our fits we use the various measured values of $g_{V,A}^{\nu e}$ and
combinations of $C_{iq}$ taken from the particle data book \cite{PDG}. For
clarity these are listed in Table~\ref{tab:lowenergy}.

\begin{table}
\begin{center}
\begin{tabular}{cccc}
\hline
\hline
Quantity & Experimental value & SM prediction & Pull [$\sigma$]\\
\hline
$g_L^2$ & $0.3010 \pm 0.0015$ & 0.3037 & -1.8\\
$g_R^2$ & $0.0308 \pm 0.0011$ & 0.0300 & 0.7\\
$\theta_L$ & $2.51 \pm 0.033$ & 2.46 & 1.4\\
$\theta_R$ & $4.59 \mbox{}^{+0.41}_{-0.23}$ & 5.18 & -1.4\\
\hline
$g_{V}^{\nu e}$ & $-0.040 \pm 0.015$ & -0.039 & -0.1\\
$g_{A}^{\nu e}$ & $-0.507 \pm 0.014$ & -0.506 & -0.1\\
\hline
$C_{1u} + C_{1d}$ & $0.147 \pm 0.004$ & 0.153 & -1.5\\
$C_{1u} - C_{1d}$ & $-0.604 \pm 0.066$ & -0.530 & -1.1\\
$C_{2u} + C_{2d}$ & $0.72 \pm 0.89$ & -0.0095& 0.8\\
$C_{2u} - C_{2d}$ & $-0.071 \pm 0.044$ & -0.062 & -0.2\\
\hline
\hline
\end{tabular}
\end{center}
\caption{\sl Results for the various model independent parameters
   which describe low energy neutral current processes (taken from
   \cite{PDG}). The value for $g_L^2$, measured by the NuTeV collaboration 
has
   been modified to take into account the strange quark asymmetry
   \cite{erler}. In fits we also include the correlations which are 
provided in
   \cite{PDG}. SM predictions are produced by GAPP using the input 
parameters
   listed in table~\ref{tab:gapp}.}
\label{tab:lowenergy}
\end{table}

\subsection{$Z$-pole}

The corrections (\ref{eq:smz}) and (\ref{eq:dg}) lead to different predictions
for the set of observables measured in $e^+ e^-$ collisions at the
$Z$-resonance. Final data from the combination of the LEP-1 and SLC results is
provided in reference \cite{ewwg}.

Several $Z$-pole observables are associated with the various $Z$ partial
widths given by 
\begin{equation}
\label{eq:Zwidth}
\Gamma (Z\to f\bar{f}) \equiv \Gamma_Z^f = \frac{e_0^2}{24\pi\,s_0^2 c_0^2}
\mph{Z} 
\left(g^{f\,2}_L + g^{f\,2}_R\right)\,,
\end{equation}
at tree-level when fermion masses are neglected. We can write
Eq.(\ref{eq:Zwidth}) in terms of the usual SM prediction plus a linearised
correction solely due to $\delta g^f_{L,R}$.
\begin{equation}
\label{eq:Zcorrections}
  \Gamma_Z^f = \Gamma_Z^{f,{\rm SM}} \left[1+\frac{2\left(g_L^{f,\rm SM}\,\delta
        g_L^f + g_R^{f,\rm SM}\,\delta
        g_R^f \right)}{\left(g_L^{f,\rm SM}\right)^2 + \left(g_R^{f,\rm SM}\right)^2} 
\right]\,.
\end{equation}
The predictions for the observables $R_{e,\mu,\tau}$, $R_{c,b}$ and
$\sigma^{\rm peak}_f$, which are all defined in Appendix 
\ref{app:zpoleobs}, can now be
written down using Eqs.(\ref{eq:dg}) and (\ref{eq:Zcorrections}).

Other $Z$-pole observables are defined from various asymmetries in the
cross-sections for $e^+ e^- \to f \bar{f}$ measured at the Z resonance. These
asymmetries are defined in Appendix~\ref{app:zpoleobs} in terms of the
parameter ${\cal A}_f$ which is defined in Eq.~\eqref{eq:af}.

Just as for the partial $Z$ widths, ${\cal A}_f$ can be expanded in terms
of a SM prediction plus a linearised correction due to $\delta g^f_{L,R}$
\begin{equation}
\label{eq:Afcorr}
  {\cal A}_f = {\cal A}_f^{\rm SM} + \frac{4 g_L^{f,\rm SM} g_R^{f,\rm SM}}{\left[\left
(g_L^{f,\rm
        SM}\right)^2 + \left(g_R^{f,\rm
        SM}\right)^2\right]^2} \left( g_R^{f,\rm SM} \delta g^f_L - g_L^{f,\rm SM} \delta 
g^f_R\right)\,.
\end{equation}

We also include the $W$ mass, $\mph{W}$, in the fit to the Z-pole data. The
expression for $\mph{W}$ in Eq.~\eqref{eq:mw} can be expressed
in terms of a $M^{\rm SM}_W$ defined from the measured input parameters $
(e,s_0)$ and the corrections due to the LWSM as follows
\begin{equation}
\mphq{W} = \msm{W}^2 (1 + \delta_W) =( M^{\rm SM}_W)^2 \frac{1 + \delta_W}
{1+ \delta_{s_0}}\,,
\end{equation}
with 
\begin{eqnarray}
M^{\rm SM}_W = \frac{e v}{2 s_0 } \,, \quad \delta_W = \frac{2}{\tilde x + \sqrt{\tilde x^2 - 4}} \,, \;\; \tilde  
x \equiv \Big( \frac{x_2}{c^2} -2 \Big) \, , \nonumber  
\\ s^2 = s_0^2 ( 1 + \delta_{s_0}) \,, \qquad 
\delta_{s_0} = \frac{c^2}{c^2-s^2}\delta_Z \,,
\end{eqnarray}
where $x_2$ is defined in Eq.~\eqref{eq:x12} 
and the relation in the last line is the
linear approximation of Eq.~\eqref{eq:famous}
with \eqref{eq:dadG}.

For clarity, in Table~\ref{tab:lep1} we list the set of LEP-1 and SLC
observables used in our fit to the Z-pole data. We use the results obtained by
assuming lepton universality. ${\cal A}_{b,c}$ are measured from
left-right-forward-backward asymmetries, c.f. \eqref{eq:FBLR} and \eqref{eq:asymm},
at SLC and $A_{\tau}^{\rm (pol)}$ is a
combination of ${\cal A}_e$ and ${\cal A}_\tau$,
c.f. \eqref{eq:taupol} \eqref{eq:tauFB} and \eqref{eq:asymm}, measured using the tau
polarisation at LEP. ${\cal A}_e(A^f_{LR})$ is a combination of measurements of ${\cal
  A}_{e,\mu,\tau}$ at SLC, which are found to be consistent with lepton
universality \cite{ewwg}. The average is dominated by the result from hadronic final states, e.g. \cite{PDG}.

\begin{table}
\begin{center}
\begin{tabular}{cccc}
\hline
\hline
Quantity & Experimental value & SM prediction & Pull [$\sigma$]\\
\hline
$\Gamma_Z$ [GeV] & $2.4952 \pm 0.0023$ & 2.4956 & -0.2\\
$\sigma_f^{\rm peak}$ [nb] & $41.540 \pm 0.037$ & 41.476 & 1.7\\
$R_{\ell}$ & $20.767 \pm 0.025$ & 20.744 & 0.9\\
$R_b$ & $0.21629 \pm 0.00066$ & 0.21580 & 0.7\\
$R_c$ & $0.1721 \pm 0.0030$ & 0.1723 & -0.1\\
$A^{(\tau)}_{pol}$ & $0.1465 \pm 0.0033$ & 0.1463 & 0.0\\
${\cal A}_e (A^f_{LR})$ & $0.1513 \pm 0.0021$ & 0.1463 & 2.4\\
${\cal A}_b$ & $0.923 \pm 0.020$ & 0.935& -0.6\\
${\cal A}_c$ & $0.670 \pm 0.027$ & 0.667 & 0.1\\
$A^{\ell}_{FB}$ & $0.0171 \pm 0.0010$ & 0.0161 & 1.0\\
$A^b_{FB}$ & $0.0992 \pm 0.0016$ & 0.1026 & -2.1\\
$A^c_{FB}$ & $0.0707 \pm 0.0035$ & 0.0733 & -0.7\\
\hline
$\mph{W}$& $80.398 \pm 0.025$ & 80.364 & 1.4\\
\hline
\hline
\end{tabular}
\end{center}
\caption{\sl Data collected at the Z resonance by LEP and SLC, taken from
   \cite{ewwg}. In the numerical fitting we also use the correlations 
provided
   by \cite{ewwg}. The latest $W$ mass combination is also included in our 
fit
   where we use the March 2008 Electroweak Working Group combined result
   \cite{wmassfit}. SM predictions are produced by GAPP using the input 
parameters listed in table~\ref{tab:gapp}.}
\label{tab:lep1}
\end{table}

\subsection{LEP-2}
As discussed in section~\ref{sec:LEP2}, we include constraints from LEP-2 data by making use of the formalism of oblique corrections. Among the 
seven oblique parameters only three $X$,$W$ and $Y$ are relevant since
the $S$, $T$ and $U$ can be exchanged into the three Altarelli \& Barbieri
parameters $\varepsilon_{1,2,3}$ \cite{ItalySpain} which are already constrained to be small
from LEP-1/SLC and the variable parameter $V$ is not relevant for
$Z$ and $\gamma$ exchanges measured at LEP-2.
Numerically, we use
constraints on the $X$,$Y$ and $W$ parameters provided in \cite{ItalySpain} which
we repeat here for completeness,
\begin{eqnarray}
X & = & (-2.3 \pm 3.5) \cdot  10^{-3}\,,\nonumber\\
Y & = & (+4.2 \pm 4.9) \cdot 10^{-3}\,,\nonumber\\
W & = & (-2.7 \pm 2.0) \cdot 10^{-3}\,,
\end{eqnarray}
with correlation matrix
\begin{equation}
\rho = \left(\begin{array}{ccc}
1 & -0.96 & +0.84 \\
-0.96 & 1 & -0.92 \\
+0.84 & -0.92 & 1\end{array}\right)\,.
\end{equation}
Using the results from Eq.~(\ref{eq:LWPI}), notice that at tree level in the
LWSM we have $X=0$ (due to the symmetry properties of the operators
added in the LWSM). The $W$ and $Y$ parameters are however non-zero and are
given by Eq.~(\ref{eq:WY}).

\subsection{Numerical Results}

\begin{figure}[!ht]
\centering
{\bf(a)}~~~\begin{minipage}{10cm}\includegraphics[width=10cm]{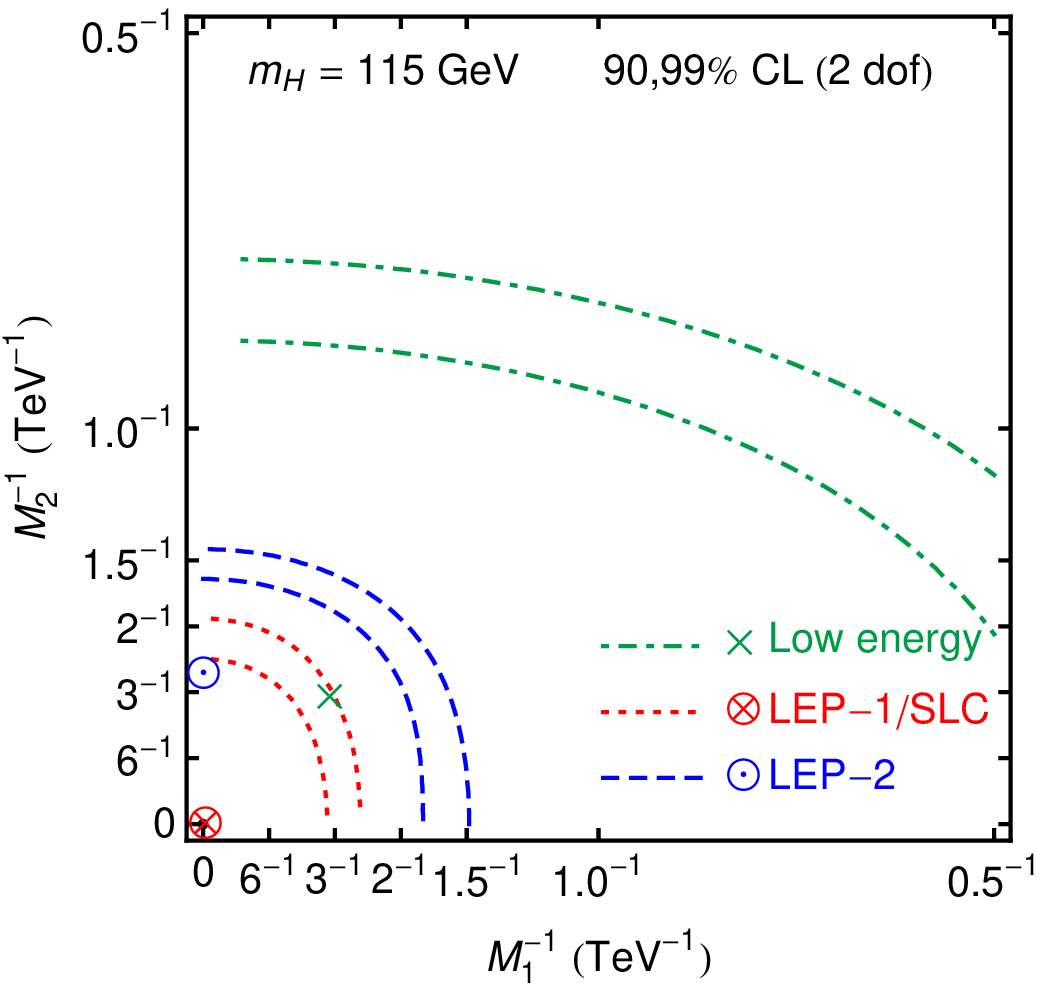}\end{minipage}
\vspace{5mm}
{\bf(b)}~~~\begin{minipage}{10cm}\includegraphics[width=10cm]{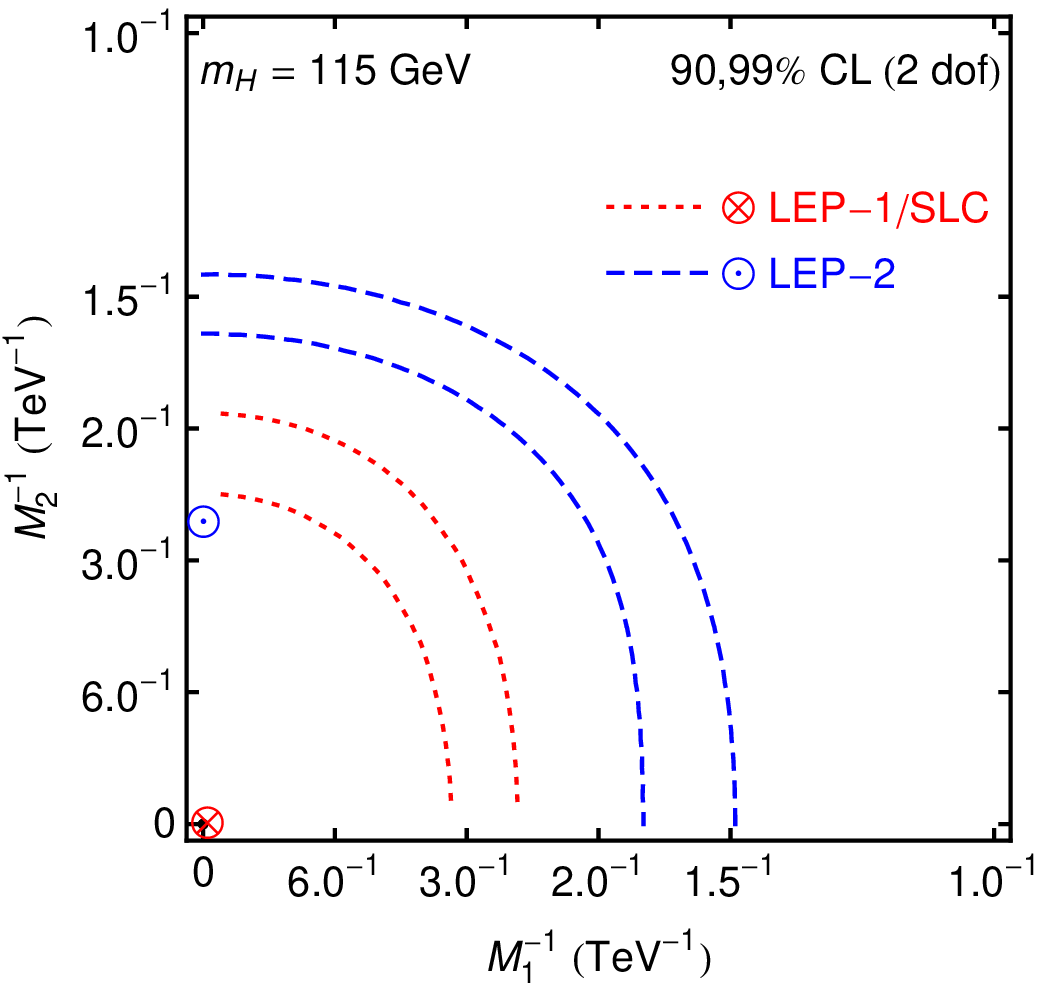}\end{minipage}
\caption{\small\sl {\bf (a)} The $1/M_1$ vs. $1/M_2$ plane showing the results of
  $2$ parameter $\chi^2$ fits to each dataset (low energy, Z-pole and LEP-2).
  The SM limit is the bottom-left corner. $90$ and $99\%$ C.L. exclusion contours
  (2 dof) are shown. {\bf (b)} The $1/M_1$ vs. $1/M_2$ plane showing only the more
  constraining Z-pole and LEP-2 fits.}
\label{fig:all}
\end{figure}

Using the results of the preceeding sections, we perform various 2-parameter
$\chi^2$ fits of the data to the LWSM by varying the LW masses $M_1$ and
$M_2$. In Figure~\ref{fig:all} we show the $90\%$ and $99\%$ C.L. exclusion contours
(2 dof) for individual fits to each dataset. We plot the contours on the
$1/M_1$ vs.  $1/M_2$ plane which has the SM limit ($M_1\to \infty$ and $M_2
\to \infty$) at a single point in the bottom left corner. The best fit points
are also marked for each dataset.

For the low energy data, the minimum $\chi^2$ lies away from the SM, with a
marginally lower $\chi^2$ such that $\chi^2/{\rm dof} = 11.2/(10-2) = 1.4$,
compared to the SM which has $\chi^2/{\rm dof} = 11.3/10 = 1.1$. This is not
the case for the much more sensitive $Z$-pole data which have a minimum
$\chi^2$ located at the point corresponding to the SM.

Figure~\ref{fig:all} clearly shows that the low energy data more tightly
constrain $M_2$ than $M_1$. The reason for this can be seen in
Eq.~(\ref{eq:formofds}), where the correction to $s^2_*(0)$ is more
sensitive to $x_2 \sim 1/M_2$ than $x_1 \sim 1/M_1$ by a factor of $c^2$
versus $s^2$.

The $\chi^2$ fit to the Z-pole data could have been performed indirectly by
using the constraints on the oblique parameters $W$ and $Y$ which come from
Z-pole data whilst the other oblique parameters are set to zero. We have
checked that virtually identical results can be obtained in this way by using
the numerical constraints provided in
\cite{ItalySpain}\footnote{\footnotesize To obtain complete agreement between
  the different approaches one must use the same set of inputs to generate the
  SM predictions. We show plots for input parameters which differ slightly
  from those used in \cite{ItalySpain}, for example we use the recently
  updated average of the top quark mass uncertainty\cite{top}}.

\begin{figure}[!ht]
\centering
{\bf(a)}~~~\begin{minipage}{10cm}\includegraphics[width=10cm]{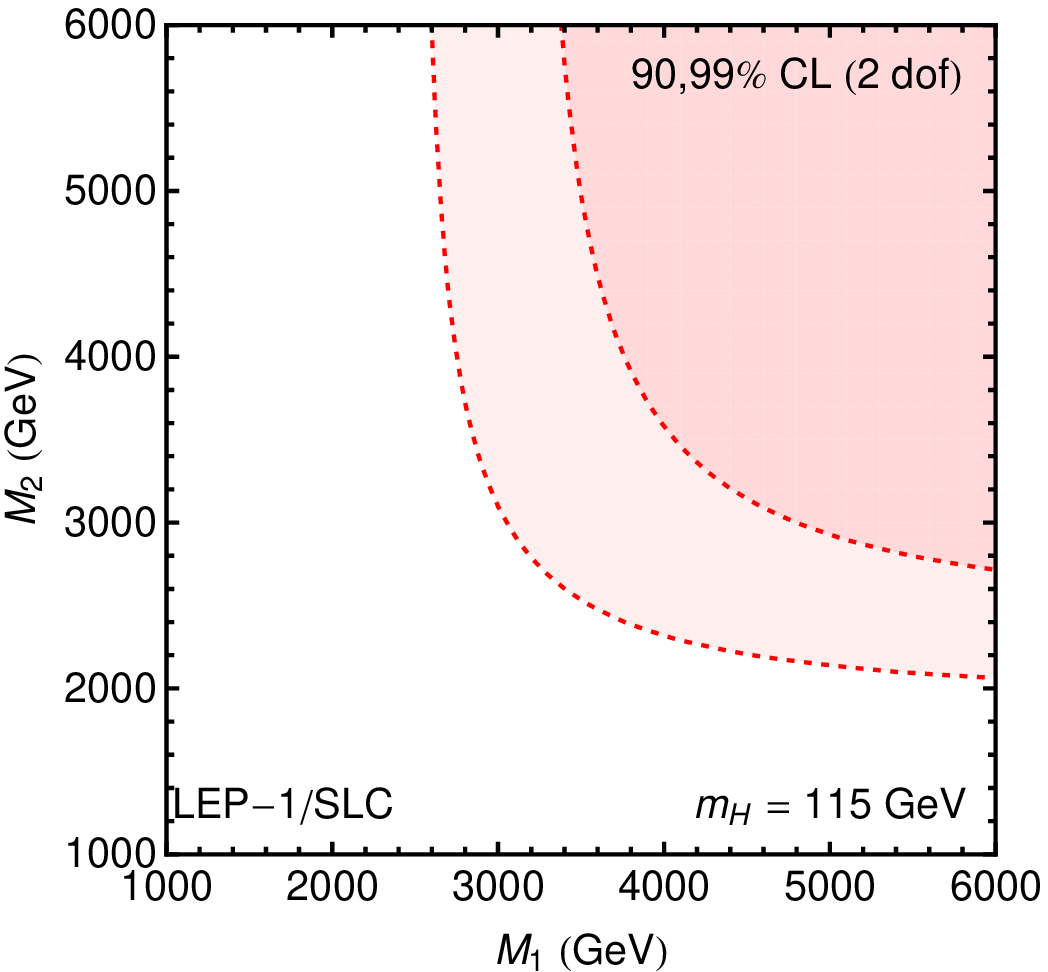}\end{minipage}
\vspace{5mm}
{\bf(b)}~~~\begin{minipage}{10cm}\includegraphics[width=10cm]{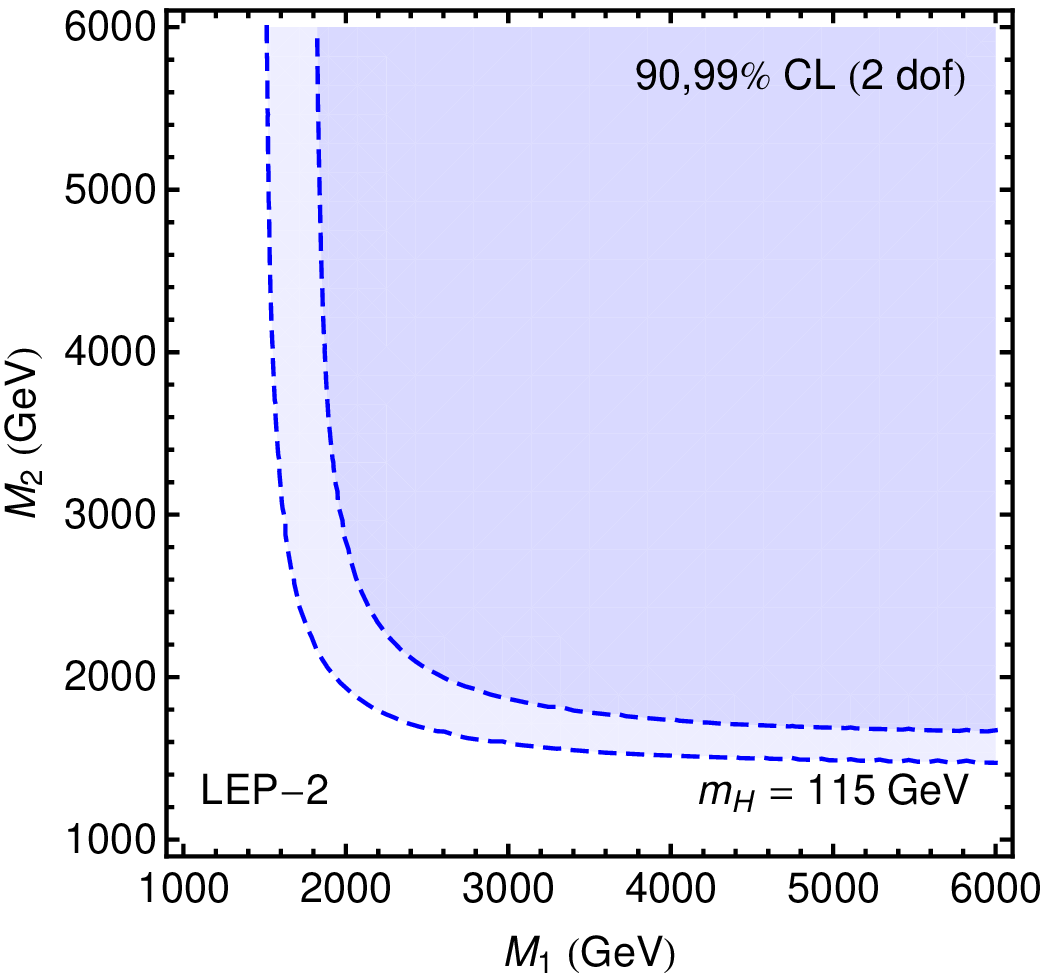}\end{minipage}
\caption{\small\sl The $M_1$ vs. $M_2$ plane showing the results of the 2
  parameter $\chi^2$ fits to the {\bf (a)} Z-pole and {\bf (b)} LEP-2 data. $90$ and $99\%$ C.L.
  exclusion contours (2 dof) are shown in each case.}
\label{fig:masses}
\end{figure}

From Figure~\ref{fig:all} we clearly see that the LEP-2 data provide less
stringent constraints on the LW masses than the $Z$-pole data\footnote{\footnotesize As the constraints on $W$ and $Y$ are taken from
  \cite{ItalySpain} in which the SM input parameters differ slightly to ours,
  the curves should not strictly be compared. However we have verified that,
  for the same input parameters, the LEP-2 data are less constraining than the
  $Z$-pole results.}. They show a minimum away from the SM corresponding to
$M_1\to \infty$ and $M_2 \simeq 2.6$~TeV.

In Figure~\ref{fig:masses} we invert these plots to show the $90$ and $99\%$
C.L.  excluded regions on the $M_1$ vs. $M_2$ plane. The constraints from the
low energy data are not shown as it is clear from Figure~\ref{fig:all} that
these data only very weakly constrain the model. The tightest constraints come
from the $Z$-pole data, however they still potentially allow the LW mass $M_2$ to
be as low as $M_1 \simeq 2$~TeV if the other LW mass is $M_1 \simge 6$~TeV.
For the case $M_1 \simeq M_2$ the mass scale $M_1 \simeq M_2 \simle 3$~TeV
is excluded at 99\% C.L. by the $Z$-pole data alone.

As the LW masses are lowered, the observables which are most problematic to
the fit are the left-right asymmetry $A^e_{LR}$ and the $W$ mass. At $M_1 =
M_2 = 3$~TeV these observables produce pulls of 3.5$\sigma$ and 2.4$\sigma$
respectively. It is interesting to note that these observables also induced
sizable pulls in the SM fit. In fact, they are averages of several individual
measurements and in the case of the $W$-mass, the measurements from lepton and
hadron colliders are only just consistent \cite{wmassfit} \footnote{A strategy to 
improve on the measurement of $W$
observables, by varying the beam energy, at the 
LHC has been put forward in reference 
\cite{wite}.}. 
It is a curious
issue that if only the LEP-2 determination was used then the $W$-mass would 
fit better in both the SM and the LW extension. There exists a similar
situation for the ${\cal A}_e (A_{LR}^f)$ which is an average (assuming lepton
universality) of several results obtained at the SLC. The result for ${\cal
  A}_e$ obtained from hadronic final states is quite large when compared to
the determinations of ${\cal A}_{\mu,\tau}$ and determination of ${\cal A}_e$
from the measurement of $A^{(\tau)}_{\rm pol}$ \cite{ewwg}. It furthermore has
quite a small error, the effect of which is to pull the average for ${\cal A}_e (A_{LR}^f)$
away from the SM prediction, and consequently makes the fit to the LWSM worse.

In light of this information, it is tempting to remove the problematic data
from the fit. For example, if the $\chi^2$ fit is performed without both the TeVatron
$W$-mass determination and the${\cal A}_e (A_{LR}^f)$ from SLC then the bounds on the LW
masses are weakened. The new fit has a minimum away from the SM, at
approximately $M_1 \simeq 3.3$~TeV and $M_2 \simeq 8.3$~TeV. The 90 and 99$\%$
C.L. exclusion contours then look rather similar to the LEP-2 constraints and
we find that $M_1 = M_2 \simeq 2.1$~GeV is not excluded at 99$\%$ C.L.

\section{Conclusions}
\label{sec:con}

We have analysed the constraints on the LWSM coming from electroweak precision data.
We have derived effective electroweak Lagrangians adequate
for low energy and the Z-pole (LEP-1/SLC) measurements to all orders in the LW masses
at tree level. We have assessed the fit to the LEP-2 data within the
oblique approximation. The only non-zero oblique parameters
are $Y = - \msm{W}^2/M_1^2$ and  $W = - \msm{W}^2/M_2^2$ [Eq.~\eqref{eq:WY}].  
 All other oblique parameters,  $(\hat S,X)$ and $(\hat T,
\hat U,V)$, are 
found to be zero at the tree level. 
Our work differs from
 previous \cite{EW1}  and later work \cite{lebed}
 by including effects of contact terms. We find that the latter are of leading order and therefore indispensable as shown in a separate addendum.
We have uncovered the negative sign of $W$ and $Y$ as a consequence
of the ghost nature of the model, c.f. subsection \ref{sec:sign}.
In subsection \ref{sec:reder} we have rederived the parameters in the low energy and Z-pole
Lagrangians in terms of the oblique parameters in the leading
$\msm{W}/M_{1,2}^2$ approximation.

By performing a $\chi^2$ fit we have produced $90\%$ and $99\%$ C.L. exclusion plots for $M_1$ and $M_2$ which are shown in
Fig.~\ref{fig:all} (Left). The low energy constraints are considerably
weaker than the ones from LEP-1/SLC and LEP-2.
Degenerate LW masses $M_1 = M_2 \simeq 3 \, {\rm TeV}$
are excluded at the $99\%$ confidence level.
Somewhat lower values $M_2(M_1) \simeq 2(2.5) \, {\rm TeV}$
are possible when one of the mass scales assumes a very large value.

Studies of $Z'$ models  \cite{AnneSylivie}
would suggest that the resonance like structures of the
LW degrees of freedom could be seen at the LHC for masses of
up to $\sim 5 \, {\rm TeV}$,  for an integrated luminosity of $100 \,
{\rm fb}^{-1}$.
On the other hand, the relatively high bounds on the masses point towards the
little hierarchy problem, which expresses 
the dilemma 
that electroweak data requires typically a light Higgs
and sets strong bounds on new degrees of freedom,
which were themselves introduced to cure the hierarchy problem.
We would like to point out the similarity between the LWSM and models with gauge bosons propagating in a flat extra
dimension \cite{ItalySpain}.  Just as in the LWSM, these models do not violate the 
first two symmetry classes $(+,-)$ and 
$(-,-)$, described in section \ref{sec:LEP2}, 
and therefore the only possible non-vanishing oblique parameters at tree level
are $W = Y = ( g_2 v \pi R)^2/6 = 2/3 \pi^2 \msm{W}^2/M_{KK}^2$
(where $R$ is the radius of the extra dimension and $M_{KK}$ is the mass of the first KK mode).
The difference between the LWSM and this situation is that $W,Y$ are positive rather than negative and this inverts the role of LEP-2 and LEP-1/SLC
data in terms of their constraining role,
as can  be inferred from the  $(W,Y)$ plot
in reference \cite{ItalySpain}.

The question of
whether quantum field theories of the LW type are consistent or not is interesting independently of their phenomenological aspects.
Does the contour deformation \cite{S-gang} lead to
a unitary perturbation theory? Is this eventually at the cost of Lorentz invariance \cite{Nakanishi}?
Does microscopic acausality not lead to macroscopic paradoxes?
\cite{Coleman-Erice}.
It is very encouraging that
these questions have a positive answers within perturbation theory in the $O(N)$ model in the large $N$ limit \cite{O(N)}.
On to other hand, one might also speculate as to whether
the LWSM is only an effective description of a theory --- potentially with effects which can already be felt by the electroweak precision data
investigated in this paper.

\section*{Acknowledgments}

We are grateful to Alexander Merle for discussions and participation at early
stages of this project. Moreover we acknowledge discussions with Oliver Brein
and Georg Weiglein on aspects of electroweak physics and Frank Krauss for
discussions on dijets and careful reading of the manuscript.  We thank Jens
Erler and John Terning for helpful correspondence. We further acknowledge
correspondence with Richard Lebed \& Christopher Carone and Ezequiel Alvarez,
Carlos Schat, Alejandro Szynkman \& Leandro Da Rold on their work.  RZ is
supported in part by the Marie Curie research training networks contract Nos.\
MRTN-CT-2006-035482, {\sc Flavianet}, and MRTN-CT-2006-035505, {\sc Heptools}.

\section*{Addendum}

After this work appeared, a further article, \cite{lebed} was submitted to
arXiv agreeing with the approach of reference \cite{EW1}.  The results of these
works differ numerically and conceptually from ours
since non-negligible contact terms are omitted.

Before entering into greater detail we would like to emphasize that our
results for the oblique parameters, which rely on an expansion in
$\msm{W,Z}^2/M_{\rm LW}^2$, are backed-up by an exact treatment of the
tree-level low-energy and $Z$-pole effective Lagrangians i.e. our results
have also been obtained completely independently of any oblique formalism.

The well-known oblique formalism, as described by Peskin \& Takeuchi
\cite{PT}, is suitable for constraining so-called universal models, where all
new physics contributions to precision measurements can be described by
modifications of the SM gauge boson self energies.  The LWSM in the auxiliary
field formalism does not belong to this class since there are extra weak gauge
bosons of the LW type.  In recent years it was realized that theories of this
kind can be brought into a universal form if these additional weak gauge
bosons couple to the SM fermions in terms of the usual $J_Y^\mu$ and $J^\mu_a$
only. This is achieved by integrating out the linear combination of the SM and
additional gauge bosons that does not couple to $J_Y^\mu$ and $J^\mu_a$, thus
avoiding the generation of contact terms (${\cal L}^{\rm eff} \sim
J_{Y,a}^2$).

In cases like this the new physics can be described by 4 leading
parameters\footnote{ Carets are used to distinguish the Barberi {\it et al.}
  $\hat{S}$ and $\hat{T}$ parameters from the older $S$ and $T$ parameters.},
$\hat S, \hat T, W, Y$, plus 3 sub-leading parameters, $X, \hat U, V$. The
additional parameters can be seen at the price for taking the contact terms
consistently into account. For a concise review on this subject, with many
working examples, we refer the reader to Barbieri {\it et al.}
\cite{ItalySpain}.

The relation between the formalism presented in Barbieri {\it et al}
\cite{ItalySpain} and the on-shell formalism used by Peskin \& Takeuchi (plus
possible contact terms) is worked out in a transparent manner in \cite{Chiv},
from where we reproduce the following formulae:

\begin{eqnarray}
\hat{S} & = & \frac{1}{4s^2}\left( \alpha S + 4 c^2 \left(\Delta \rho_*(0) - \alpha
    T\right) + \frac{\alpha \delta}{c^2}\right)\,,\\
\hat{T} & = & \Delta \rho_*(0)\,,\\
\label{W}
W & = & \frac{\alpha \delta}{4 s^2 c^2}\,,\\
\label{Y}
Y & = & \frac{c^2}{s^2}\left(\Delta \rho_*(0) - \alpha T\right)\,,
\end{eqnarray}
or inverted,
\begin{eqnarray}
\alpha S & = & 4 s^2\left(\hat{S}-Y-W\right)\,,\label{alphaS}\\
\alpha T & = & \hat{T} - \frac{s^2}{c^2} Y\,,\label{alphaT}\\
\alpha \delta & = & 4 s^2 c^2 W\,,\label{adelta}\\
\Delta \rho_*(0) & = & \hat{T}\label{aDrho}\,, 
\end{eqnarray}
with $\Delta \rho_*(0) \equiv \rho_*(0) -1$, c.f.  Eq.~\eqref{eq:para} in our
notation.  The quantities $\delta \sim W$ and $(\Delta \rho - \alpha T) \sim
Y$ parameterize the effects of the charged and neutral current contact terms;
c.f formulae (2.9) and (2.10) in reference \cite{Chiv}.

In references \cite{EW1,lebed} only the LW gauge bosons were integrated out
and it was assumed that the contact terms are negligible. This is explicitly
stated in section III of reference \cite{EW1}.  In reference \cite{lebed} the
same effective action is obtained, from where identical results to \cite{EW1}
follow\footnote{Note that in reference \cite{lebed} the special case of $M_1
  \to \infty$ is considered and $S$ and $T$ are presented in that limit.
  Nevertheless it is stated that they agree with \cite{EW1} when $M_1$ is kept
  finite.}.

Using the results obtained in this paper and Eqs. \eqref{W} and \eqref{Y} it
is apparent that the contact terms are of leading order and are therefore not
negligible.  This can be seen for instance from the rho parameter $\rho_*(0) =
1 + \Delta \rho_*(0)$ which is equal to one in our paper with $\hat T = 0$
\eqref{eq:Class2}, but with the un-modified Peskin \& Takeuchi formalism
$\rho_*(0) = 1 + \alpha T$ [c.f. \cite{PT} Eq.~(3.13)] it apparently receives
a leading order correction since $T \sim m_W^2/M_{LW}^2$ as we shall see
below.

Besides low energy physics such as $\rho_*(0)$, the contact terms also affect
Z-pole physics in an indirect way through a shift in $G_F$ which in turn
modifies the extraction of the Weinberg angle, e.g.  in Eq.~\eqref{eq:celebre}
or Eq.~(3.4) in \cite{PT}.
 
In the LWSM, using the formulae \eqref{alphaS} and \eqref{alphaT}, the
relations between $S$ and $T$ in the Peskin \& Takeuchi formalism and $\hat
S,\hat T, W,Y$ in the Barbieri {\it et al} formalism can be illustrated,
allowing a cross-check of the calculations. It is found that
\begin{eqnarray}
S\Big|_{\mbox{\small Eq.~\eqref{alphaS}}}  &=& \frac{4 s^2}{\alpha} \left( \frac{m_W^2}{M_1^2}
+ \frac{m_W^2}{M_2^2}\right) = 4 \pi v^2
 \left( \frac{1}{M_1^2}
+ \frac{1}{M_2^2}\right)\,,  \\[0.1cm]
\label{eq:T}
T\Big|_{\mbox{\small Eq.~\eqref{alphaT}}} 
 &=& \frac{s^2}{c^2 \alpha}  \frac{m_W^2}{M_1^2}
= \pi v^2 \frac{g_1^2+g_2^2}{g_1^2} \frac{1}{M_1^2} \,,
\end{eqnarray}
which does indeed agree with the results of reference \cite{EW1} Eqs. (7) and (8)
and reference \cite{lebed} in the limit $M_1 \to \infty$.


The crucial point, and thus the problem with the results of references
\cite{EW1} and \cite{lebed}, is that the constraints on $S$ and $T$ from, for
example, the LEP Electroweak Working Group will be produced assuming that the
4-fermion contact terms, parameterized by $W$ \eqref{W} and $Y$ \eqref{Y}, are
absent or negligible. The important point is that the relations between $S$
and $T$ and the experimental observables will be modified by the non-zero $W$
\eqref{W} and $Y$ \eqref{Y} parameters [or equivalently the non-zero values of
$\delta$, Eq.~\eqref{adelta}, and ($\Delta \rho_*(0) - \alpha T$)].
Constraining the LWSM with $S$ and $T$ {\it a la} Peskin \& Takeuchi is
therefore unsuitable, and the cause of our differences with references
\cite{EW1} and \cite{lebed}.

\appendix
\setcounter{equation}{0}
\renewcommand{\theequation}{A.\arabic{equation}}

\section{$W,Z,A$ propagators in the HD formalism}
\label{app:WZprop}

The LW $W$-propagator in the HD picture can be derived by taking into
consideration the additional kinetic term of the gauge field
\begin{equation}
\delta {\cal L} = \frac{1}{M_2^2} {\rm Tr} [ (\hat D \hat W)_\mu^2]\,,
\end{equation}
and the Higgs field
\begin{equation}
\label{eq:HiggsHD}
\delta {\cal L} =  - \frac{1}{M_H^2} \hat D^2 \hat H(\hat D^2 \hat H)^\dagger \, .
\end{equation}
We also introduce a standard $R_\xi$ gauge fixing term\footnote{The gauge fixing 
term is actually not needed
as long as the Higgs VEV is non zero. Omitting it simply corresponds to the unitary 
gauge $\xi \to \infty$.}
\begin{equation}
{\cal L}^{gf} = -\frac{1}{2 \xi } (\partial \! \cdot \! \hat A)^2\,.
\end{equation}
The propagator can then be derived in a straightforward way and is eventually given by
\begin{equation}
\hat D^W_{\mu\nu} = \frac{i}{p^2 -\msm{W}^2 -p^4/M_2^2}\Big( - g_{\mu\nu} + p_
{\mu} p_\nu f^W_{pp} \Big)\,,
\end{equation}
with
\begin{equation}
f^W_{pp} = \frac{M_2^2(( 1- \xi)M_H^2 + \xi \msm{W}^2) + M_H^2 p^2 }{M_2^2 
M_H^2 p^2 + \msm{W}^2 M_2^2 (p^2 -M_H^2)\xi } \, ,
\end{equation}
where $\msm{W}$ is defined in Eq.~\eqref{eq:mwsm}. The propagator reduces to the LW 
gauge field propagator as given in Eq.~(24) of reference \cite{GOW}
in the limit,
\begin{equation}
\lim_{\msm{W}^2 \to 0}  f^W_{pp} =  \frac{1-\xi}{p^2} + \frac{\xi}{M_2^2} \,,
\end{equation}
where the Higgs VEV vanishes. Moreover, in the LW decoupling limit,
\begin{eqnarray}
\lim_{M_H^2,M_2^2 \to \infty}  f^W_{pp} &=& \frac{(1-\xi)}{p^2 - \xi \msm{W}^2}\,,
\end{eqnarray}
the propagator reduces to the standard SM $W$-propagator. 

For $M_1 = M_2$, the $Z$-propagator is the same as the $W$-propagator with
the simple replacement $\msm{W} \to \msm{Z} \equiv (g_2 v/2)/c$. If $M_1 \neq 
M_2$ then
the $(\hat W_3, \hat B)$ system cannot be simultaneously diagonalised. 
The propagator in the $(\hat Z,\hat A)$-basis  is given by 
\begin{equation}
\label{eq:DN}
\hat D_{\mu \nu}^{N}(p^2) = \hat D^{N}(p^2) \left(- g_{\mu \nu} + p_\mu p_\nu f^{N}_
{pp} \right)\,,
\end{equation}
where
\begin{eqnarray}
\label{eq:DNexp}
&  & \hat D^{N}(p^2) =  i(\Gamma^{-1}) \,, \qquad   f_{pp}^N =  \mathbb{I}_2 - 
( \mathbb{I}_2 -X)^{-1}\,, \nonumber \\[0.1cm] 
&  & \Gamma = p^2  \mathbb{I}_2 - m_0^2 - p^4 R_W^T  M^{-2} R_W\,, \nonumber 
\\[0.1cm]
&  & X = \left[  \mathbb{I}_2 - \Gamma^{-1} m_0^2 \left(1 -\frac{p^2}{M_H^2}\right) \right]
- \frac{1}{\xi}\left[  \mathbb{I}_2 - \Gamma^{-1} \left(\Gamma -  \mathbb{I}_2 p^2 \right) \right]\,,
\end{eqnarray}
with 
\begin{equation}
m_0^2 =  \left( \begin{array}{cc}
\msm{Z}^2 \hspace*{5mm} & 0 \\
0 \hspace*{5mm} & 0 \end{array} \right)\,,\qquad 
M^{-2} =  \left( \begin{array}{cc}
1/M_2^2  \hspace*{5mm} & 0 \\
0 \hspace*{5mm} & 1/M_1^2 \end{array} \right)\,,
\end{equation}
and $R_W$ defined as in Eq.~\eqref{eq:weinrot}.
We will not give the explicit form of the matrix $\Gamma^{-1}$ but only the relevant low energy limits
\begin{eqnarray}
\label{eq:G1}
& \lim_{p^2 \to 0}& (-i) \left(\hat D^N\right)_{ZZ} = - \frac{1}{\msm{Z}^2}\,, \\
\label{eq:G2}
& \lim_{p^2 \to 0}& (-i) \left(\hat D^N\right)_{AA}-\frac{1}{p^2} = \left(\frac{c^2}{M_1^2} + \frac
{s^2}{M_2^2}\right)\,,\\
\label{eq:G3}
& \lim_{p^2 \to 0}&  (-i) \left(\hat D^N\right)_{ZA} =  0 \,.
\end{eqnarray}
Note that the photon pole was explicitly subtracted from Eq.~\eqref{eq:G2}.
The $(Z,Z)$ component Eq.~\eqref{eq:G1} is the same as in the SM, whereas the $
(A,A)$ component has
the expected additional massive particle contribution in addition to the photon which contributes 
to 
the low energy effective Lagrangian. The non-diagonal contribution Eq.~\eqref{eq:G3} 
vanishes  in the low energy limit.
All terms in the non-diagonal part are of course proportional to $M_1^2 - M_2^2$.

We have given the formal expression for the matrix  $f_{pp}^{N}$ related to the 
$p_\mu p_\nu$ structure only for completeness. The low energy limits are 
smooth and do not affect the order ${\cal O}(m_{f} m_{f'}/
\mphq{W})$ supression.

\section{The degenerate case $M_1 = M_2$}
\label{app:m12eq}
\setcounter{equation}{0}
\renewcommand{\theequation}{B.\arabic{equation}}

When the LW masses, $M_2$ and $M_1$, of the SU(2)$_L$ and U(1)  gauge fields are equal
matters become analytically tractable.

\subsection{Mass matrix diagonalisation}

The heavy neutral gauge boson mass matrix, denoted by
\begin{equation}
{\cal M_B}^{\prime\prime\prime\prime}= 
\left( \begin{array}{cccc}
 \mphq{Z} & 0&  0 \\
0 & -  \mphq{\tilde A} & 0 \\
0 & 0 & -  \mphq{\tilde Z} \\
\end{array}\right) \,,
\end{equation}
can be diagonalised by the transformation defined in Eq.~\eqref{eq:Q}
with
\begin{equation}
\label{eq:follow}
 \phi = 0 \quad  \Rightarrow  \quad x \equiv \frac{M^2}{\msm{Z}^2} =  \left(1+ \Tanh
{\theta}\right)\left(1+ \frac{1}{\Tanh{\theta}}\right)\,,
\end{equation}
with masses 
\begin{alignat}{1}
\label{eq:EVeq}
\frac{ \mphq{Z}}{\msm{Z}^2} &  =
1+ \Tanh{\theta} =
\frac{1}{2}\left(x - \sqrt{x(x-4)}\right)\,,  \nonumber \\
\frac{ \mphq{\tilde A}}{\msm{Z}^2} & \,=\,  \left(1+ \Tanh{\theta}\right)\left(1+ \frac{1}{\Tanh
{\theta}}\right) =
x\,,  \nonumber \\
\frac{ \mphq{\tilde Z}}{\msm{Z}^2} &= 
1+ \frac{1}{\Tanh{\theta}} =
\frac{1}{2}\left(x + \sqrt{x(x-4)}\right)  \, .
\end{alignat}
We have denoted $x \equiv x_1 = x_2$ in the degenerate limit. Its analytical
expression is the $\phi \to 0$ limit of the expression given in Eq.~\eqref{eq:x12}. 
The value for $M_Z^2$ in Eq.~\eqref{eq:EVeq} is of course the limit
of the general expression for $M_Z^2$ given in Eq.~\eqref{eq:smz}.
The combined transformation $S \cdot Q$ (c.f. Eqs.\eqref{eq:weinrot} and \eqref{eq:Q}) is used 
to parameterize the diagonalization  and $\phi = 0$ if and only if $M_1 = M_2$ which is 
due to the fact that in this case the transformation
$$
S' = R_W \otimes \mathbb{I}_2\,,
$$
decouples or diagonalizes the photon and the LW photon immediately.
This leaves behind a reduced $2\times 2$ matrix which is then 
diagonalized by the single hyperbolic rotation angle $\theta$. 
This system is then algebraically identical to the $W$ system discussed in Chapter 
2.4.2 in \cite{KUZ} for instance.

\subsection{The Z-pole effective Lagrangian in the HD formalism}
\label{app:HDM}
The purpose of this subsection is to derive the Z-pole effective Lagrangian (the same as in section \ref{sec:zpoleHD}) in a simpified and therefore
more transparent setup.
For the case $M_1 = M_2$ the higher deriavtive propagator can be 
diagonalized in the $(\hat Z,\hat A)$ space c.f. appendix \ref{app:WZprop}.
The couplings 
of the gauge boson to the currents are left unchanged up to corrections
of order ${\cal O}(m_f^2/M_f^2)$. The Weinberg angle keeps its original meaning 
and this leads to the prediction
$$
s_*^2(\mph{Z}) = s^2 \,,
$$
which is indeed verified in the limit $\phi \to 0$ c.f. Eq.~\eqref{eq:smz}.
We will uncover the parameter $\rho_f$ as a multiplicative factor
to the standard propagator. 
The scalar HD propagator \cite{KUZ}  reads 
\begin{equation}
\label{eq:prop}
\hat D(p^2) = \frac{i}{p^2 - p^4/M^2 - m^2} = \frac{-i M^2}{(p^2 - m_{\rm ph}^2)(p^2 - M_{\rm ph}^2 )}\, .
\end{equation}
The analogue of the matching equation \eqref{eq:trick} for $\rho_f$
in the case $M_1 = M_2$ then looks like 
\begin{equation}
  \lim_{p^2 \to \mphq{Z}} (-i) \hat D(p^2)\left(p^2 - \mphq{Z}\right) =
\rho_f  \, \underbrace{\left(4 \mphq{Z} \sqrt{2}G_F\right)\frac{c^2}{g_2^2} }_{1+\delta_Z}\,.
\end{equation}
This leads to 
\begin{eqnarray}
\label{eq:hm}
\rho_f (1+\delta_Z)  &=& \frac{-M^2}{ \mphq{Z} -  \mphq{\tilde Z}} \ =\ 
\frac{ \mphq{\tilde Z} +  \mphq{Z}}{  \mphq{\tilde Z}- \mphq{Z} } \ =\ \frac{x}{\sqrt{x
(x-4)}}  
\nonumber \\
&=&   \frac{2 + \Tanh{\theta} +\Tanh{\theta}^{-1}}{\Tanh{\theta} -\Tanh{\theta}^{-1}} \ =\ 
\cosh^2(\theta) (1+ \Tanh{\theta})^2 \,,
\end{eqnarray}
where we have used  used Eq. \eqref{eq:EVeq} in transforming it to its final
form. This leads to 
\begin{equation}
\label{eq:rhofeq}
\rho_f = \cosh^2(\theta) (1 +  \Tanh{\theta})\,,
\end{equation}
in accordance with Eq.\eqref{eq:smz} in the limit $\phi \to 0$.
The factor $\rho_f (1+\delta_Z)$ is in fact the scaling factor 
due to hyperbolic rotations. In reference  \cite{KUZ} we have
denoted it by $s_{(A-\tilde A)^2}$ and indeed
\begin{equation}
s_{(Z-\tilde Z)^2} \equiv  \frac{1 + r_Z^2}{1-r_Z^2}  = \rho_f (1+\delta_Z) \,,
\end{equation}
with the notation  $r_Z \equiv  \mph{Z}/ \mph{\tilde Z}$ and using Eq.~\eqref
{eq:hm}.
In the simplified case $M_1 = M_2$ discussed here it is most transparent
that $\rho_f \geq 1$ in Eq.~\eqref{eq:rhofeq} since it is easily deduced from 
Eq.~\eqref{eq:EVeq} that $1+\Tanh{\theta}$ takes values
in the range $[2,1[$ when $x$ is constrained to be in its allowed region  $[4,\infty[$.

\section{Z pole observables}
\label{app:zpoleobs}
\setcounter{equation}{0}
\renewcommand{\theequation}{C.\arabic{equation}}

In this appendix we define the Z-pole observables used
throughout section \ref{sec:fit} to constrain the mass scales of the 
LW electroweak gauge boson masses.
The Z-pole observables are defined from the decay rates of the Z boson and various 
asymmetries in the cross section for $e^+ e^- \to \bar f f$ at a 
centre of mass energy equal to the mass of the $Z$ boson.

\subsection{Ratios of decay rates}
The ratios $R_f$, which are a direct extension of the famous $R$ function to the Z 
pole, are defined from the decay rates
\begin{equation}
\Gamma_Z^f \equiv \Gamma( Z  \to \bar f f) \quad {\rm and} \quad \Gamma_Z^{\rm 
had} = 
\sum_ {f = u,d,s,c,b} \Gamma( Z  \to \bar f f)  \, ,
\end{equation} 
where the tree level expression for the rate has been given in
Eq.~\eqref{eq:Zwidth}.
The leptonic ratios $R_{e,\mu,\tau}$ and the quark ratios $R_{c,b}$ are given
by 
\begin{equation}
\label{eq:R}
R_{e,\mu,\tau} = \frac{\Gamma_Z^{\rm had}}{\Gamma_Z^{e,\mu,\tau}}\,, 
\qquad R_{c,b} = \frac{\Gamma_Z^{c,b}}{\Gamma_Z^{\rm had}} \,.
\end{equation}
The total Z decay rate is given as the sum of the leptonic and hadronic rate
\begin{equation}
\Gamma_Z = \Gamma_Z^l + \Gamma_Z^{\rm had} + 3 \Gamma_Z^{\bar \nu \nu} \, 
.
\end{equation}
\subsection{Cross-sections \& Asymmetries}
\subsubsection{General defintions}
Introducing the following shorthand notation for the cross section for $e^+e^- \to 
\bar f f$
\begin{equation}
\sigma_f \equiv \sigma(e^+e^- \to \bar f f)\, ,
\end{equation}
the peak cross section is simply the cross section at a centre of mass energy equal to the Z boson mass
\begin{equation}
\sigma_f^{\rm peak} = \sigma_f(s = \mphq{Z}) \,.
\end{equation}

It is possible to define five types of asymmetries by combining cross sections for events in the forward and backward hemispheres, and events with differing initial and final 
state polarizations in all possible ways.
The forward backward asymmetry, 
\begin{equation}
A_{FB}^f \equiv \frac{\sigma_F^f -\sigma_B^f}{\sigma_F^f +\sigma_B^f} \,,
\end{equation}
is defined as the normalized difference of
the cross-sections for the electron-like fermion 
going into the forward and backward hemispheres.

The polarization of  the initial state electron is used to define 
the left-right asymmetry,
\begin{equation}
\label{eq:LR}
A_{\rm LR} \equiv  \frac{\sigma_L^f -\sigma_R^f}{\sigma_L^f +\sigma_R^f} \,,
\end{equation}
as the normalized difference of the cross-sections for scattering with left and
right handed electrons. 
These two asymmetries can also be combined into,
\begin{equation}
\label{eq:FBLR}
A_{\rm FBLR} \equiv  \frac{(\sigma_F-\sigma_B)_L^f -(\sigma_F-\sigma_B)_R^f}
{(\sigma_F-\sigma_B)_L^f +(\sigma_F-\sigma_B)_R^f} \,,
\end{equation}
the forward backward left-right asymmetry.
We have assumed maximal polarization in Eqs.~\eqref{eq:LR} and
\eqref{eq:FBLR} c.f. for example \cite{ewwg} Eqs. (1.58) and (1.59) for more
details.

If the polarization of the final state can be measured, as it was possible at
the SLAC large detector (SLD) with $\tau$ leptons, then an analoguous asymmetry can be defined for 
them as well. 
The final state left right asymmetry is defined
\begin{equation}
\label{eq:taupol}
A_{rl}^\tau \equiv A_{\rm pol}^{(\tau)} \equiv  \langle P_\tau \rangle \equiv
 \frac{\sigma_r^\tau -\sigma_l^\tau}{\sigma_r^\tau +\sigma_r^\tau}\,,
\end{equation}
and the forward backward asymmetry is defined
\begin{equation}
\label{eq:tauFB}
A_{\rm FB}^{\rm pol} \equiv  \frac{(\sigma_r-\sigma_l)_F^\tau -(\sigma_r-\sigma_l)
_B^\tau}{(\sigma_r-\sigma_l)_F^\tau +(\sigma_r-\sigma_l)_B^\tau}\,.
\end{equation}

\subsection{Results at the Z-pole}

At the Z-pole the asymmetries and the peak cross sections assume
a very simple form, since we can neglect photon-photon and Z-photon exchange 
terms. The peak cross section is simply given by
\begin{equation}
\sigma_f^{\rm peak} = \frac{12 \pi}{\mphq{Z}} \frac{\Gamma_Z^e \Gamma_Z^f}
{\Gamma_Z^2} \,. 
\end{equation}
Working at tree level and with zero fermion masses the
observables are
\begin{eqnarray}
\label{eq:asymm}
A_{\rm FB}^f &=& \frac{3}{4} {\cal A}_e {\cal A}_f \nonumber \\
A_{\rm LR}^f  &=&  {\cal A}_e \nonumber \\
A_{\rm FBLR}^f  &=&  \frac{3}{4}{\cal A}_f \nonumber \\
\langle P_\tau \rangle   &=& -  {\cal A}_\tau \nonumber \\
A_{\rm FB}^{\rm pol} &=& - \frac{3}{4}  {\cal A}_e \,,
\end{eqnarray}
with the notation,
\begin{equation}
\label{eq:af}
{\cal A}_f \equiv  \frac{(g^f_L)^2 - (g^f_R)^2}{(g^f_L)^2 + (g^f_R)^2} 
\,.
\end{equation}

\subsection{Low energy observables}
\label{app:low}

Low energy electroweak processes particularly sensitive to new physics are $\nu$-nucleon and $\nu$-electron scattering and parity violating electron-hadron interactions
 parameterized by the following effective Lagrangians \cite{PDG}
 \begin{eqnarray}
 \label{eq:lowEW}
 {\cal L}^{\nu N} &=& -\frac{G_F}{\sqrt{2}} (\bar \nu \nu)_{V\!-\!A} 
\sum_{q = u,d} \big( \epsilon_L(q) (\bar q q )_{V\!-\!A}  +  
 \epsilon_R(q) (\bar q q )_{V\!+\!A} \big)\,,\nonumber \\[0.1cm]
 {\cal L}^{\nu_\mu e} &=& -\frac{G_F}{\sqrt{2}} (\bar \nu \nu)_{V\!-\!A} 
\big( g_{V}^{\nu e} (\bar e e )_{V}  - 
g_{A}^{\nu e} (\bar e e )_{A} \big)\,,\nonumber \\[0.1cm]
 {\cal L}^{e q} &=& \frac{G_F}{\sqrt{2}} \sum_{q = u,d} \big(
 C_{1q} (\bar e e)_A (\bar q q)_V + C_{2q} (\bar e e)_V (\bar q q)_A
 \big)\,.
 \end{eqnarray}
The parameters of the deep inelastic neutrino scattering Lagrangian
 $\epsilon_{L(R)}(u(d))$ are determined from ratios of cross 
 sections such as the Llewellyn-Smith ratio, $R_\nu$ or the
 Paschos Wolfenstein ratio, $R^-$ \cite{PDG}. However, instead of the $\epsilon_{L(R)}(u(d))$, we will use the following equivalent but less correlated set parameters
 \begin{equation}
 g_L^2 \equiv \epsilon_L^2(u) + \epsilon_L^2(d) \,, \qquad 
 g_R^2 \equiv \epsilon_R^2(u) + \epsilon_R^2(d) \,,
 \end{equation}
and the isospin breaking parameters 
 \begin{equation}
 \label{eq:iso}
 \theta_{L(R)} \equiv \tan^{-1}\Big(\frac{\epsilon_{L(R)(u)}}{\epsilon_{L(R)(d)}}\Big) 
\,,
 \end{equation}
obtained from fits to the data. Information on the parameters $\theta_{L(R)}$ can 
 be obtained by varying the isoscalarity of the nucleon target.
  
The parameters $g_{V(A)}^{e \nu}$ are obtained from
$\nu_\mu e \to \nu_\mu e$ scattering \cite{PDG} which 
is transmitted by a t-channel $Z$ boson at leading order.

There are many types of experiment  determining the
various linear combinations of the coefficients $C_{(1,2)(u,d)}$  
such as polarization experiments of $e_{L(R)} N \to e X$, $e_{L(R)} D \to e X$
or measuring the admixture of a P wave contribution to the 
6s ground state of Cesium \cite{PDG}.

\end{document}